%% file: main.tex
\documentclass[review=false]{acmart}

\makeatletter
\def\@ACM@checkaffil{
    \if@ACM@instpresent\else
    \ClassWarningNoLine{\@classname}{No institution present for an affiliation}%
    \fi
    \if@ACM@citypresent\else
    \ClassWarningNoLine{\@classname}{No city present for an affiliation}%
    \fi
    \if@ACM@countrypresent\else
        \ClassWarningNoLine{\@classname}{No country present for an affiliation}%
    \fi
}
\makeatother

\copyrightyear{2024}
\acmYear{2024}
\setcopyright{acmlicensed}\acmConference[CHI '24]{Proceedings of the CHI Conference on Human Factors in Computing Systems}{May 11--16, 2024}{Honolulu, HI, USA}
\acmBooktitle{Proceedings of the CHI Conference on Human Factors in Computing Systems (CHI '24), May 11--16, 2024, Honolulu, HI, USA}
\acmDOI{10.1145/3613904.3642884}
\acmISBN{979-8-4007-0330-0/24/05}

\AtBeginDocument{%
  \providecommand\BibTeX{{%
    \normalfont B\kern-0.5em{\scshape i\kern-0.25em b}\kern-0.8em\TeX}}}

\def\markup{0}

\if\markup1
\usepackage[normalem]{ulem} 
\newcommand{\rv}[1]{{\leavevmode\color{blue}#1}}

\else
\newcommand{\rv}[1]{#1}
\newcommand{\st}[1]{}
\newcommand{\sout}[1]{}
\fi

\graphicspath{{Figures/}}

\usepackage{appendix}
\usepackage{float}
\usepackage{hyperref}
\usepackage{multirow} 
\usepackage{makecell}
\usepackage{rotating}
\usepackage{subcaption}
\usepackage{verbatim}
\usepackage{lineno}

\begin{document}
\nolinenumbers
\title{Designing Upper-Body Gesture Interaction with and for People with Spinal Muscular Atrophy in VR}



\author{Jingze Tian}
\authornote{Both authors contributed equally to this research.}
\affiliation{%
  \institution{Southeast University}
  \country{China}
}
\affiliation{%
  \institution{The Hong Kong University of Science and Technology (Guangzhou)}
  \country{China}
}
\email{tjz@seu.edu.cn}

\author{Yingna Wang}
\authornotemark[1]
\affiliation{%
  \institution{The Hong Kong University of Science and Technology (Guangzhou)}
  \country{China}
}
\email{ywang885@connect.hkust-gz.edu.cn}

\author{Keye Yu}
\affiliation{%
  \institution{University College London}
  \country{UK}}
\affiliation{%
  \institution{The Hong Kong University of Science and Technology (Guangzhou)}
  \country{China}
}
\email{zczlky4@ucl.ac.uk}

\author{Liyi Xu}
\affiliation{%
  \institution{Nanjing University of the Arts}
  \country{China}
}
\affiliation{%
  \institution{The Hong Kong University of Science and Technology (Guangzhou)}
  \country{China}
}
\email{xuliyi1@nua.edu.cn}

\author{Junan Xie}
\affiliation{%
  \institution{The Hong Kong University of Science and Technology (Guangzhou)}
  \country{China}
}
\email{jxie622@connect.hkust-gz.edu.cn}
 
\author{Franklin Mingzhe Li}
\affiliation{%
  \institution{Carnegie Mellon University}
  \country{USA}
}
\email{mingzhe2@cs.cmu.edu}

\author{Yafeng Niu}
\affiliation{%
  \institution{Southeast University}
  \country{China}
}
\email{nyf@seu.edu.cn}

\author{Mingming Fan}
\authornote{Corresponding author}
\affiliation{%
  \institution{The Hong Kong University of Science and Technology (Guangzhou)}
  \country{China}
}
\affiliation{%
  \institution{The Hong Kong University of Science and Technology}
  \country{China}}
  
\email{mingmingfan@ust.hk}

\begin{abstract}

    Recent research proposed gaze-assisted gestures to enhance interaction within virtual reality (VR), providing opportunities for people with motor impairments to experience VR. Compared to people with other motor impairments, those with Spinal Muscular Atrophy (SMA) exhibit enhanced distal limb mobility, providing them with more design space. However, it remains unknown what gaze-assisted upper-body gestures people with SMA would want and be able to perform. We conducted an elicitation study in which 12 VR-experienced people with SMA designed upper-body gestures for 26 VR commands, and collected 312 user-defined gestures. Participants predominantly favored creating gestures with their hands. The type of tasks and participants' abilities influence their choice of body parts for gesture design. Participants tended to enhance their body involvement and preferred gestures that required minimal physical effort, and were aesthetically pleasing. Our research will contribute to creating better gesture-based input methods for people with motor impairments to interact with VR.

\end{abstract}




\begin{teaserfigure}
  \includegraphics[width=\textwidth]{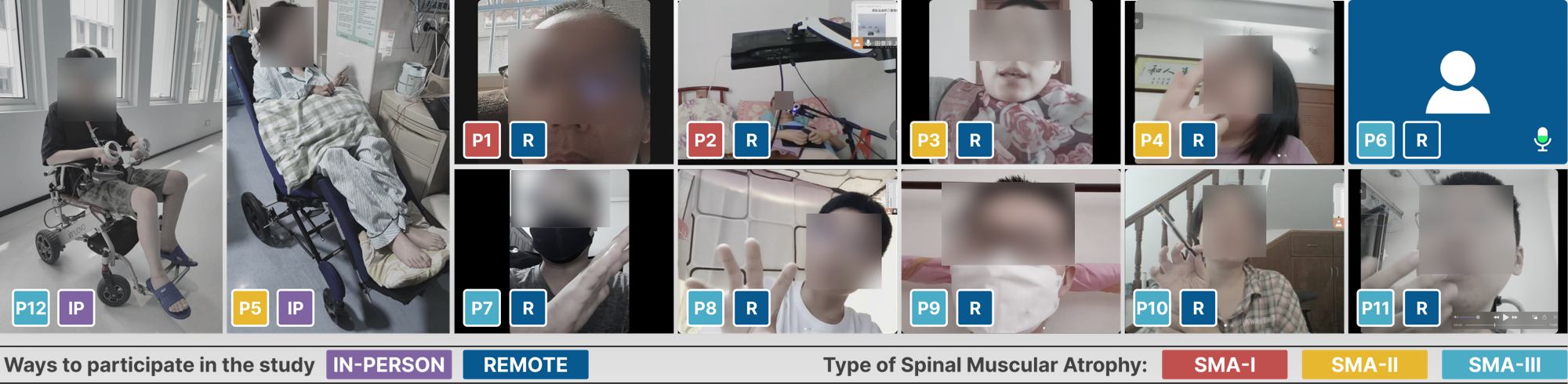}
  \caption{\textbf{Twelve people with Spinal Muscular Atrophy designed upper-body gestures for VR commands. Two of them participated in person, while the other ten joined remotely. All participants actively took part in this research, and Participant 6 preferred to conduct the experiments without turning on the camera.}}
  \label{fig:teaser}
  \Description{Figure 1 shows twelve people with spinal muscular atrophy (SMA) designing upper-body gestures for VR commands. Two of them participated in person, while the other ten joined remotely. All participants actively took part in this research, although Participant 6 preferred to conduct the experiments without turning on the camera. The figure shows the screenshots of the user study, and it illustrates the process of participants performing gestures with various parts of their upper bodies, including their eyes, mouth, hands, arms, and shoulders. }
\end{teaserfigure}


\maketitle

\section{Introduction}
\input{Introduction}

\section{RELATED WORK}
\input{Related_Work}

\section{METHOD}
\input{Method}

\section{Results}
\input{Results}

\section{Discussion}
\input{Discussion}
 
\section{Limitation and Future Work}
\input{Limitation_and_Future_Work}

\section{Conclusion}
\input{Conclusion}

\begin{acks}
\begin{sloppypar}
This work is partially supported by 1) 2024 Guangzhou Science and Technology Program City-University Joint Funding Project (PI: Mingming Fan); 2) 2023 Guangzhou Science and Technology Program City-University Joint Funding Project (Project No. 2023A03J0001); 3) Guangdong Provincial Key Lab of Integrated Communication, Sensing and Computation for Ubiquitous Internet of Things (No.2023B1212010007).
\end{sloppypar}
\end{acks}

\bibliographystyle{ACM-Reference-Format}
\bibliography{main}


\appendix

\section{The reference for the six VR Categories}
\label{app:VR Categories}

In Section 3.2, We selected six VR categories candidates from the referenced paper \cite{spittleReviewInteractionTechniques2023}, or official documents of Oculus or HTC Vive, as shown in Table \ref{table: VR Categories Reference}. We initially finalized six VR categories candidates, as shown in Table \ref{table:VR Categories Candidates}. It is noted that \textit{Pointing} and \textit{Viewport Control} need to receive continuous real-time signals from the VR device, they are not well-suited for transforming into commands for user-defined gestures. Therefore, we finally gathered four VR Commands Category presented in Table \ref{table: Fundamental Categories of VR Commands}.

\begin{table*}[ht!]
    \caption{\textbf{Categories of VR Commands and their References. Each VR Commands category is supported by at least one reference, sourced from the referenced paper, or official documents of Oculus or HTC Vive.}}
    \Description{Categories of VR Commands and their References. Each VR Commands category is supported by at least one reference, sourced from the referenced paper, or official documents of Oculus or HTC Vive. This table presents various categories of VR commands, descriptions of each category, and how they are referred to in Oculus and HTC Vive documentation.}
    \label{table: VR Categories Reference}
    \centering
    \resizebox{\textwidth}{!}{%
    \begin{tabular}{ll|lll}
    \hline
    \multicolumn{2}{c|}{\textbf{Our Work}} &
      \multicolumn{3}{c}{\textbf{Reference}} \\ \hline
    \multicolumn{1}{c|}{\textbf{\begin{tabular}[c]{@{}c@{}}VR Categories\\ Candidate\end{tabular}}} &
      \multicolumn{1}{c|}{\textbf{Description}} &
      \multicolumn{1}{l|}{\textbf{\begin{tabular}[c]{@{}l@{}} A review of interaction\\ techniques for immersive\\ environments \cite{spittleReviewInteractionTechniques2023}\end{tabular}}} &
      \multicolumn{1}{c|}{\textbf{Oculus Document}} &
      \multicolumn{1}{c}{\textbf{HTC Vive Document}} \\ \hline
    \multicolumn{1}{l|}{\textbf{Pointing}} &
      \begin{tabular}[c]{@{}l@{}}The act of locating interactive \\ elements through methods such \\ as virtual hand hovering or ray \\ casting.\end{tabular} &
      \multicolumn{1}{l|}{Pointing} &
      \multicolumn{1}{l|}{\begin{tabular}[c]{@{}l@{}}Pointer\\ \\ Cursor\end{tabular}} &
      Content Targeting \\ \hline
    \multicolumn{1}{l|}{\textbf{Selection}} &
      \begin{tabular}[c]{@{}l@{}}Initiating or confirming an \\ action after pointing, such \\ as grabbing an object up \\ close or from a distance.\end{tabular} &
      \multicolumn{1}{l|}{Selection} &
      \multicolumn{1}{l|}{Select Something} &
      \begin{tabular}[c]{@{}l@{}}Selection:\\ \\ Grab and Place\end{tabular} \\ \hline
    \multicolumn{1}{l|}{\textbf{Manipulation}} &
      \begin{tabular}[c]{@{}l@{}}Moving, rotating, or resizing \\ interactive elements, as well \\ as altering their properties.\end{tabular} &
      \multicolumn{1}{l|}{\begin{tabular}[c]{@{}l@{}}Translation\\      \\ Rotation\\ \\ Scaling\end{tabular}} &
      \multicolumn{1}{l|}{\begin{tabular}[c]{@{}l@{}}Move Something\\ \\ \\ Rotate Something\\ \\ \\ Resize Something\end{tabular}} &
      Manipulation \\ \hline
    \multicolumn{1}{l|}{\textbf{Viewport Control}} &
      \begin{tabular}[c]{@{}l@{}}Zooming and panning within an \\ environment using dedicated \\ functions.\end{tabular} &
      \multicolumn{1}{l|}{Viewport Control} &
      \multicolumn{1}{l|}{} &
      Targeting \\ \hline
    \multicolumn{1}{l|}{\textbf{\begin{tabular}[c]{@{}l@{}}Menu-Based\\ Interaction\end{tabular}}} &
      \begin{tabular}[c]{@{}l@{}}Presenting a structured set \\ of tabs, commands, or \\ utilities for users to\\ engage with.\end{tabular} &
      \multicolumn{1}{l|}{Menu-based} &
      \multicolumn{1}{l|}{\begin{tabular}[c]{@{}l@{}}Buttons\\ \\ Pinch-and-Pull \\ Components\end{tabular}} &
      Context Menu \\ \hline
    \multicolumn{1}{l|}{\textbf{Locomotion}} &
      \begin{tabular}[c]{@{}l@{}}Moving or changing the \\ direction of an avatar's \\ position within a virtual \\ space.\end{tabular} &
      \multicolumn{1}{l|}{} &
      \multicolumn{1}{l|}{Locomotion} &
       \\ \hline
    \end{tabular}%
}
\end{table*}

\begin{table*}[ht]
\caption{\textbf{Six VR Categories candidates and their descriptions.}}
\Description{This table lists six candidate categories for VR gestures, provides a description for each, and indicates whether they were selected for gesture design.

Pointing:Description: The act of locating interactive elements through methods such as virtual hand hovering or ray casting.
Whether Selected as Gestures Design: No

Selection:Description: Initiating or confirming an action after pointing, such as grabbing an object up close or from a distance.
Whether Selected as Gestures Design: Yes
Manipulation:Description: Moving, rotating, or resizing interactive elements, as well as altering their properties.
Whether Selected as Gestures Design: Yes
Viewport Control:Description: Zooming and panning within an environment using dedicated functions.
Whether Selected as Gestures Design: No
Menu-Based Interaction:Description: Presenting a structured set of tabs, commands, or utilities for users to engage with.
Whether Selected as Gestures Design: Yes
Locomotion:Description: Moving or changing the direction of an avatar’s position within a virtual space.
Whether Selected as Gestures Design: Yes
The table provides a clear overview of the types of VR interactions considered and identifies which ones have been chosen for further development in gesture design.}
\label{table:VR Categories Candidates}
\centering
\resizebox{\textwidth}{!}{%
    \begin{tabular}{l|l|l}
    \hline
    \textbf{VR Categories Candidates} &
      \textbf{Description} &
      \textbf{\begin{tabular}[c]{@{}l@{}}Whether Selected\\ as Gestures Design\end{tabular}} \\ \hline
    \textbf{Pointing} &
      \begin{tabular}[c]{@{}l@{}}The act of locating interactive elements through methods\\ such as virtual hand hovering or ray casting.\end{tabular} &
      No \\ \hline
    \textbf{Selection} &
      \begin{tabular}[c]{@{}l@{}}Initiating or confirming an action after pointing, such\\ as grabbing an object up close or from a distance.\end{tabular} &
      Yes \\ \hline
    \textbf{Manipulation} &
      \begin{tabular}[c]{@{}l@{}}Moving, rotating, or resizing interactive elements, as\\ well as altering their properties.\end{tabular} &
      Yes \\ \hline
    \textbf{Viewport Control} &
      \begin{tabular}[c]{@{}l@{}}Zooming and panning within an environment using dedicated\\ functions.\end{tabular} &
      No \\ \hline
    \textbf{Menu-Based Interaction} &
      \begin{tabular}[c]{@{}l@{}}Presenting a structured set of tabs, commands, or\\ utilities for users to engage with.\end{tabular} &
      Yes \\ \hline
    \textbf{Locomotion} &
      \begin{tabular}[c]{@{}l@{}}Moving or changing the direction of an avatar's position\\ within a virtual space.\end{tabular} &
      Yes \\ \hline
    \end{tabular}%
}
\end{table*}

\section{The complete visual illustration of Hand Gestures Categories}
\begin{figure*}[ht]
    \centering
    \includegraphics[width=\linewidth]{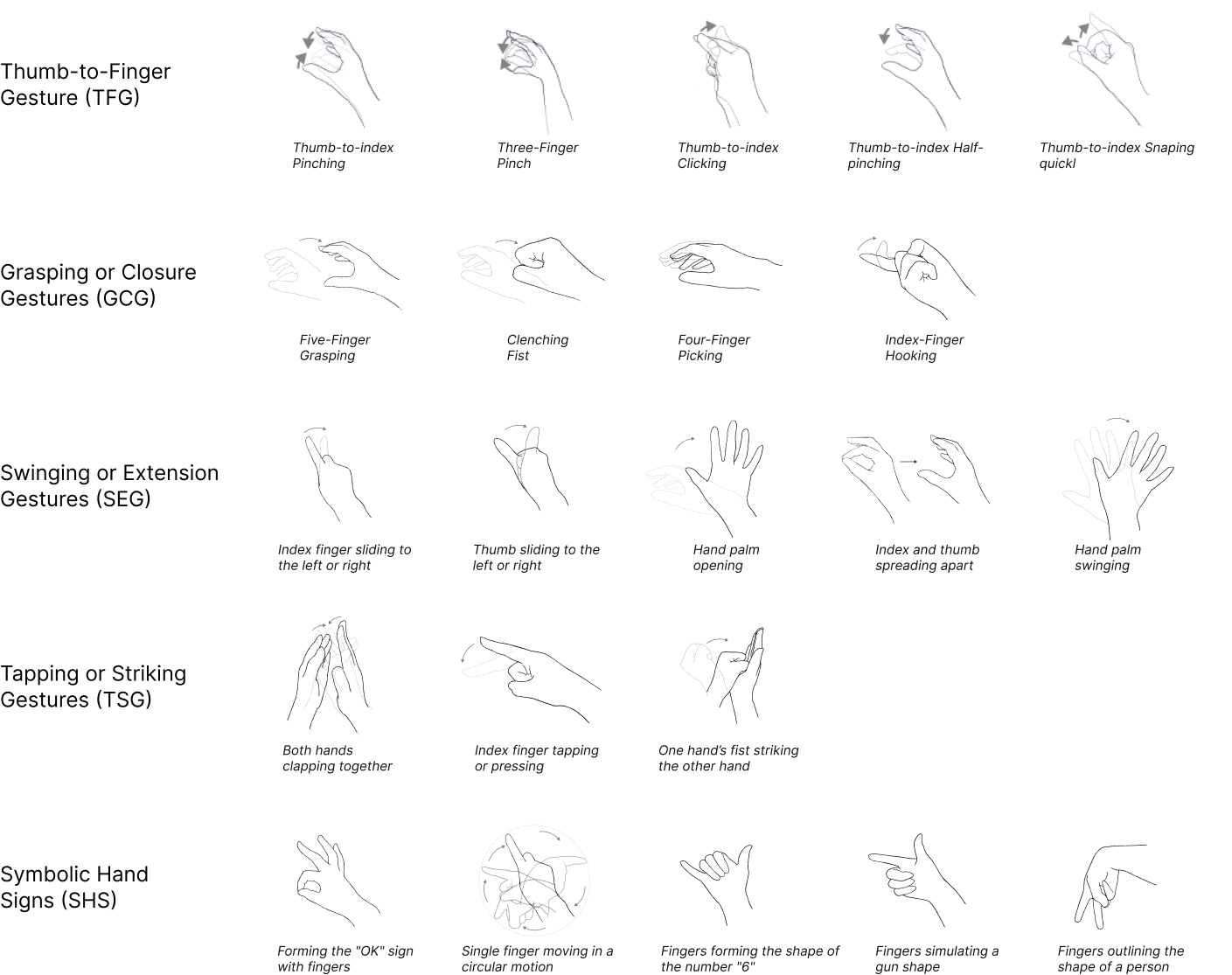}
    \caption{The complete visual illustration of Hand Gestures Categories.}
    \label{fig:The complete visual illustration of Hand Gestures Categories.}
    \Description{Figure 11 shows the complete visual illustration of Hand Gestures Categories.

The Hand Gestures Categories include five gesture sets, Thumb-to-Finger Gesture (TFG), Grasping or Closure Gestures (GCG), Swinging or Extension Gestures (SEG), Tapping or Striking Gestures (TSG) and Symbolic Hand Signs (SHS). 

The first category, Thumb-to-Finger Gesture (TFG), includes Thumb-to-index Pinching, Three-Finger Pinch, Thumb-to-index Clicking, Thumb-to-index Half-pinching and Thumb-to-index Snaping quickly. 

Grasping or Closure Gestures (GCG) includes Five-Finger Grasping, Clenching Fist, Four-Finger Picking and Index-Finger Hooking. 

Swinging or Extension Gestures (SEG) includes Index finger sliding to the left or right, Thumb sliding to the left or right, Hand palm opening, Index and thumb spreading apart and Hand palm swinging.

Tapping or Striking Gestures (TSG) includes Both hands clapping together, Index finger tapping or pressing and One hand's fist striking the other hand.

And the last set, Symbolic Hand Signs (SHS) includes Forming the "OK" sign with fingers, Single finger moving in a circular motion, Fingers forming the shape of the number "6", Fingers simulating a gun shape and Fingers outlining the shape of a person.}
\end{figure*}

The complete visual illustration of five basic hand gesture patterns, as shown in Figure \ref{fig:The complete visual illustration of Hand Gestures Categories.}
\label{app:The complete visual illustration of Hand Gestures Categories}

\end{document}

%% file: Introduction.tex
Spinal muscular atrophy (SMA) is the second most prevalent fatal autosomal recessive disorder following cystic fibrosis, afflicting approximately 1 in 6,000 to 1 in 10,000 live births \cite{oginoGeneticRiskAssessment2002,priorNewbornCarrierScreening2010}. 
This disease is characterized by generalized muscle weakness and atrophy predominating in proximal limb muscles and classified into four phenotypes (SMA I, II, III, IV) based on onset age and motor function. Unlike other motor impairments such as limb loss, cerebral palsy, or amyotrophic lateral sclerosis (ALS), SMA primarily affects central body muscles, causing difficulties in breathing, swallowing, head control, and sitting \cite{darrasSpinalMuscularAtrophies2015}. People with SMA often exhibit better distal limb mobility, such as in the hands or feet, compared to proximal limb mobility. This leads to unique interaction patterns and provides them with opportunities to utilize technologies such as VR devices. However, research on SMA's impact on human-computer interaction (HCI) is still limited. The HCI community has not fully tackled the unique challenges faced by people with SMA.

 Virtual reality (VR) can immerse users in a computer-generated environment that simulates reality both visually and interactively. VR enables people with limited mobility to engage in activities beyond their physical abilities and facilitates the exploration of inaccessible real-world experiences \cite{gerlingVirtualRealityGames2020,pei2023embodied}. Furthermore, VR promotes inclusivity and equality by granting equal mobility to all individuals \cite{lewisAssistiveTechnologyLearning1998}. 
	
 However, VR devices, which are designed with implicit \textit{Ability Assumption} \cite{kouroupetroglouAssistiveTechnologiesComputer2014a}, pose accessibility challenges for those with limited mobility, particularly due to inaccessible input methods like motion controllers and buttons \cite{mottJustWentIt2020,li2022freedom}. To utilize predominant input methods involving hand tracking or controller usage and interact with virtual objects, users need to elevate their hands to chest level, leading to prolonged muscle tension in the arms and resulting in significant fatigue. Moreover, individuals with upper-body motor impairments, particularly those with conditions like SMA, encounter greater challenges in manipulating intricate controllers and utilizing buttons that are not easily reachable. This arises from potential strength limitations that hinder their ability to access all buttons or simultaneously press and hold them. \sout{Luckily, some alternative input methods beyond motion controllers have been proposed to make VR more accessible for people with motor impairments.}

\rv{Luckily, recent research has demonstrated the effectiveness of ‘\textit{Gaze + Gestures}’, the combination of eye-tracking and hand gestures, for VR operations \cite{pfeufferGazePinchInteraction2017,slambekovaGazeGestureBased2012,luGazePinchMenuPerforming2021,poukeGazeTrackingNontouch2012}. As eye movements are inherent parts of motor planning and precede actions \cite{landWhatWaysEye2001}, this approach leverages an interactive paradigm of \textit{gaze selection} and \textit{gesture confirmation}, capitalizing on the swift nature of eye movements with gestural dexterity. People with SMA have stronger distal upper limb mobility and can proficiently use their hands, making ‘\textit{Gaze + Gestures}’ interaction a viable option. ‘\rv{\textit{Gaze + Gestures}’, has utilized in commercial products like Apple Vision Pro \footnote{\url{https://www.apple.com/apple-vision-pro/}}, holds promise as an interactive modality of VR systems for people with SMA.}}
 However, it remains unclear how they would like to use this interaction paradigm for VR interaction and modify it using other upper-body parts.
 
Our research was motivated by the need for accessible VR input methods for people with SMA and these ‘\textit{Gaze + Gestures}’ works. Our user-defined gestures design method was inspired by the prior success of designing user-defined gestures for people with motor impairments in other contexts \cite{zhaoDonWantPeople2022}. 
\sout{Motivated by this need and prior success of designing user-defined gestures in other contexts \cite{zhaoDonWantPeople2022}, } \rv{In this work,} we engaged \rv{12} people with SMA to design \textit{upper-body gestures} for \rv{26} VR \rv{common commands} and extended the design that include \textit{eye},\textit{ mouth}, \textit{face}, \textit{head}, and \textit{remained upper limb mobility} based on previous studies \cite{nukarinenEvaluationHeadTurnInteraction2016,rozadoFaceSwitchLowCostAccessibility2015,haseebHeadGesturebasedControl2018,wangIntelligentWearableVirtual2018}.
 \sout{Specifically, we conducted a user-centered study, in which 12 participants with various spinal muscular atrophy (SMA) conditions designed upper-body gestures for 26 VR commands that were commonly performed in VR systems.} During the study, participants watched video clips explaining each VR command and its effect (i.e., \textit{referent}) and then designed and performed an upper-body gesture. Afterward, participants rated the effort required to design the gesture and the mental demand, physical demand, and \rv{satisfaction level} of their created gestures using the 7-point Likert scale. Finally, we conducted semi-structured interviews to \rv{learn more about their considerations}. In total, we collected 312 user-defined gestures and identified their preference and mental models to VR input methods. 

\rv{Compared to the accessible gesture inputs for users with upper-body motor impairments for smartphone \cite{zhangSmartphoneBasedGazeGesture2017,fanEyelidGesturesMobile2020,fanEyelidGesturesPeople2022} and wearable devices \cite{vatavuUnderstandingGestureInput2022b}, our study is unique in three aspects. Firstly, our gestures are designed for VR, which is  beneficial \cite{gerlingVirtualRealityGames2020} but inaccessible \cite{mottJustWentIt2020} for people with motor impairments. Secondly, our research specifically targets people with SMA. We use user-defined upper-body gestures to explore VR input accessible for people with motor impairments, grounded in their ability and creativity.
Finally, compared to previous user-defined gestures research for people with motor impairments \cite{fanEyelidGesturesMobile2020,fanEyelidGesturesPeople2022}, we expanded the gesture design space to the upper body, which provides more freedom for them to design.}

Our paper makes the following contributions: 

        (1) We identified and described a set of 26 \rv{common} commands to cover \rv{general} VR interactions. \sout{This set was used by people with SMA to design and perform upper-body gestures in this research.}  
        
        (2) We uncovered a taxonomy of user-defined upper-body gestures based on the gestures designed by people with SMA to complete the aforementioned set of commands.
        
        (3) We derived the mental models and considerations of participants with SMA when designing upper-body gestures to increase the accessibility of VR interaction.

%% file: Related_Work.tex
Our work is informed by prior work on \textit{upper-body gestures input for people with upper-body motor impairments}, \textit{gaze-assisted gestures interaction in VR}, and \textit{user-defined gesture designs}.
 
\subsection{Upper-body Gestures Input for People with \sout{Upper-body }Motor Impairments}

\sout{Prior work has explored methods to help people with upper-body motor impairments interact with computing systems by sensing their upper-body gestures. These approaches utilize either readily available devices (e.g., smart phone \cite{zhangSmartphoneBasedGazeGesture2017,fanEyelidGesturesMobile2020,fanEyelidGesturesPeople2022} and wearables \cite{vatavuUnderstandingGestureInput2022b}), or customized technologies (e.g., \cite{rozadoFaceSwitchLowCostAccessibility2015,haseebHeadGesturebasedControl2018,ishimaruBlinkEyeCombining2014}).} Previous research has explored gesture-based interactions for individuals with upper-body motor impairments, utilizing readily available devices (e.g., smartphones \cite{zhangSmartphoneBasedGazeGesture2017,fanEyelidGesturesMobile2020,fanEyelidGesturesPeople2022} and wearables \cite{vatavuUnderstandingGestureInput2022b}) or custom technologies (e.g. \cite{rozadoFaceSwitchLowCostAccessibility2015,haseebHeadGesturebasedControl2018,li2021choose,ishimaruBlinkEyeCombining2014}). \rv{These approaches successfully transform movements into computing commands, demonstrating the feasibility and providing a diverse design space for inventing personalized gestures.}

\textit{Hand-based gestures}, such as stroke or motion gestures, are commonly used in computing device interactions. Empirical research indicates that individuals with motor impairments can accurately perform these gestures with devices worn on the wrist, finger, and head, although challenges related to repetition exist \cite{vatavuUnderstandingGestureInput2022b,vatavuStrokeGestureInputPeople2019b}. Moreover, the performance of stroke gestures for individuals with motor impairments can be enhanced through computer modeling and synthesis \cite{ungureanGestureInputUsers2018a}.

\textit{Eye-based gestures}, including eyelid and eye-gaze gestures, offer support to individuals with limited hand mobility. \textit{Eye-gaze gestures} involve relative eye movement or gaze transitions on the screen \cite{drewesInteractingComputerUsing2007,heikkilaSimpleGazeGestures2012,bigham2021fly,zhangSmartphoneBasedGazeGesture2017}. For example, Drewes et al. \cite{drewesInteractingComputerUsing2007} translated various gaze directions into media control commands. Additionally, Zhang et al. \cite{zhangSmartphoneBasedGazeGesture2017} used gaze gestures (e.g., looking up or down) for character input. \textit{Eyelid gestures} involve controlling the states and \textit{duration} of eyelids \cite{shawEyeWinkControl1990}. Heikkila et al. used prolonged eye closure as a stop command for applications. Fan et al. \cite{fanEyelidGesturesMobile2020,fanEyelidGesturesPeople2022} introduced nine eyelid gestures designed by combining different sequences, frequencies, and \textit{duration} of eye-opening and closing to control smartphones.

Gestures based on other body parts, such as the face or head, can also enhance gesture interaction for people with motor impairment. Facial features, like raising eyebrows or opening the mouth, can be utilized for customized computer control commands \cite{sun2021teethtap,rozadoFaceSwitchLowCostAccessibility2015} and to operate AR/VR headsets \cite{wangIntelligentWearableVirtual2018}. Head motions, such as turning or nodding, could be employed to control the movement of a dual-arm industrial robot \cite{haseebHeadGesturebasedControl2018} or adjust the volume of a sound system \cite{nukarinenEvaluationHeadTurnInteraction2016}.

\rv{People with SMA, a subgroup of those with motor impairments, can benefit from gesture designs based on the above research. Unlike other motor impairments (e.g., ALS, cerebral palsy, limb deficiencies), individuals with SMA often have improved fine motor skills in distal limbs, specifically in hand dexterity and finger flexibility. This enhanced ability in distal limbs presents opportunities for diverse gesture designs. Despite limited research on their gestures design, we focus on exploring upper-body gestures created by people with SMA. Moreover, we extended the gestures design space to include all the upper-body parts, including eyes, mouth, face, head, limbs, etc., to allow them to better design a richer set of user-defined gestures. }

\subsection{Gaze-assisted Gestures Interaction in VR}

\rv{In this work, we called the novel VR interaction methods that combine \textit{gaze-assisted interaction} and \textit{gesture-based interaction} as \textit{gaze-assisted gestures interaction}. \textit{Gaze-assisted interaction} involves using eye gaze to choose and visually indicate the object of interest, essentially serving as a virtual pointer in VR. The combination of gaze-based selection and gesture-based command input significantly reduces the physical effort compared to virtual hand and controller devices \cite{pfeufferGazePinchInteraction2017}, which brings hope for people with SMA to use VR devices.}

\rv{Previous studies have explored the application of \textit{gaze-assisted gestures interaction} in various VR tasks} (e.g., \textit{3D object-related interaction} \cite{pfeufferGazePinchInteraction2017,slambekovaGazeGestureBased2012,yuGazeSupported3DObject2021,sidenmarkOutlinePursuitsGazeassisted2020}, and \textit{menu-related interaction} \cite{luGazePinchMenuPerforming2021,reiterLookTurnOnehanded2022,pfeufferEmpiricalEvaluationGazeenhanced2020}) by decomposing complex VR tasks into multiple smaller sub-tasks, and build a larger design space with different integration, coordination, and transition between gaze and gestures \cite{yuGazeSupported3DObject2021}. \textit{3D object-related interaction} tasks, mainly selection and manipulation could be simplified as "\textit{gaze for selection, gesture for confirming and manipulation}" \cite{slambekovaGazeGestureBased2012}, like using \textit{‘Gaze + Pinch'} interaction \cite{pfeufferGazePinchInteraction2017}. Additionally, for \textit{menu-related interaction}, Reiter et al. \cite{reiterLookTurnOnehanded2022} used gaze to indicate menu selection, and rotational turn of the wrist to navigate the menu and manipulate continuous parameters. Lu et al. \cite{luGazePinchMenuPerforming2021} proposed a Gaze-Pinch menu with continuously performs multiple gestures on the gazing object concurrently. \rv{However, these studies were designed by people without motor impairments, it remains unknown about the \textit{gaze-assisted gestures} designed by individuals with motor challenges.}

\sout{	Prior research has shown that gaze-assisted gesture interaction requires a smaller physical effort\cite{pfeufferEmpiricalEvaluationGazeenhanced2020}, which brings hope for people with SMA to use VR devices. However, the design process of above-mentioned methods did not involve people with motor impairments. Consequently, it remains unknown how people with motor impairments could and want to use gaze-assisted gesture interaction in VR.}

\rv{We aimed to fill the existing gap by focusing on tailoring this interaction paradigm to meet the unique needs of individuals with motor impairments (i.e., people with SMA in this work) in VR. To prevent constraints on the design and encourage them to fully leverage mobility capabilities, we used the method of user-defined gestures but did not fully develop the gaze-assisted gestures, to gain insights from their preferences and challenges they may face in VR.}
\subsection{User-defined Gestures Designs}

User-defined gestures have been widely used as an elicitation study to discover and identify gesture vocabularies. It has been proved that user-defined gestures are easier to remember and learn than those defined by researchers \cite{nacentaMemorabilityPredesignedUserdefined2013}. Wobbrock et al. \cite{wobbrockUserdefinedGesturesSurface2009} started user-defined gesture for multi-touch surface computing, which was the first to employ users, rather than principles, in the development of a gesture set. They first recruited non-technical participants without prior experience using touch screens and presented the referents, or effects of an action to them, and then elicited the set of gestures meant to invoke them by using a think-aloud protocol and video analysis. This process for gesture design has been applied in a variety of domains, for example, keyboards \cite{baillyMetamorpheAugmentingHotkey2013}, public displays \cite{kurdyukovaStudyingUserdefinedIPad2012}, tangible systems \cite{valdesExploringDesignSpace2014,krayUserdefinedGesturesConnecting2010}, smartwatches \cite{arefinshimonExploringNontouchscreenGestures2016}, in-car user interfaces \cite{weidnerInteractYourCar2019}, and augmented reality \cite{piumsomboonUserdefinedGesturesAugmented2013}. As for VR, Wu et al. \cite{wuUserdefinedGestureInteraction2019} reported a research project on user-defined gestures for VR shopping applications which derived two gestures from each participant in the prior stage and selected the top-two gestures among all of them. Besides, Moran-Ledesma et al. \cite{moran-ledesmaUserDefinedGesturesPhysical2021} presented an elicitation study to manipulative gestures for 20 CAD-like and open-world game-like referents (the effect of an action in VR). Nanjappan et al. \cite{nanjappanUserelicitedDualhandInteractions2018a} presented a similar user-elicitation study for manipulating 3D objects in virtual reality environments. Our research was motivated by these user-defined methods. Specifically, our research adopts a similar user-centered approach by investigating what \textit{gaze-assisted upper-body gestures} people with Spinal muscular atrophy would like to create and how they would want to use such gestures to accomplish tasks of the VR system.

%% file: Method.tex
\subsection{Participants}

\begin{table*}[ht]
    \centering
    \caption{\textbf{Participants’ demographic information. Type indicates the severity of SMA, with I being the most severe type, followed by II and III. IP/R indicates whether the participant participated in the experiment online remotely (R) or offline in person (IP). All participants have experience using either smartphone-based, PC, or all-in-one VR headsets.}}
    \label{table:Demographic background}
    \Description{Table 1. presents the demographic information of the 12 participants in the study. It includes details of participant number, age, gender, type of spinal muscular atrophy (SMA), whether they participated remotely (R) or in person (IP), prior virtual reality (VR) experience including devices used, time length of usage, and experience content. The table shows that there were 7 males and 5 females with an average age of 28.3 years old. Two participants had SMA type I, 3 had type II, and 7 had type III. Most participants participated remotely except 2 who joined in person. All participants had prior VR experience using smartphone-based, PC, or all-in-one VR headsets, with usage time ranging from 15 minutes to over 7 hours. Their VR experience content included 360-degree movies, interactive 360-degree videos, VR games, and VR Chat.}
    \resizebox{\textwidth}{!}{
    \begin{tabular}{cccccccc}
    \hline
    \multirow{2}{*}{\textbf{Participant}} & \multirow{2}{*}{\textbf{Age}} & \multirow{2}{*}{\textbf{Sex}} & \multirow{2}{*}{\textbf{SMA type}} & \multirow{2}{*}{\textbf{IP/R}} & \multicolumn{3}{c}{\textbf{Prior VR Experience}} \\
    \cline{6-8} 
     &  &  &  &  & VR Devices & Time Length & Experience Content \\ \hline
    P1 & 28 & M & I & R & Smartphone-based VR headsets & >30min & 360° movies \\
    P2 & 25 & M & I & R & Smartphone-based VR headsets & >30min & Interactive 360° videos \\
    P3 & 26 & F & II & R & Smartphone-based VR headsets & >5h & Interactive 360° videos \\
    P4 & 26 & F & II & R & Smartphone-based VR headsets & >30min & Interactive 360° videos \\
    P5 & 20 & F & II & IP & All-in-one VR headsets & >30min & 360° movies \\
    P6 & 42 & F & III & R & All-in-one VR headsets & >15min &  Games\\
    P7 & 25 & M & III & R & PC VR headsets, Smartphone-based VR headsets & >15min & Games, 360° movies \\
    P8 & 28 & M & III & R & All-in-one VR headsets & >30min/day, >2 year & Games \\
    P9 & 20 & M & III & R & Smartphone-based VR headsets & >30min & Interactive 360° videos \\
    P10 & 26 & F & III & R & All-in-one VR headsets, Smartphone-based VR headsets & >20min & 360° movies \\
    P11 & 37 & M & III & R & All-in-one VR headsets, Smartphone-based VR headsets & >7h & 360° movies, VR Chat \\
    P12 & 37 & M & III & IP & All-in-one VR headsets, Smartphone-based VR headsets & >30min & Interactive 360° videos \\ \hline
    
    \end{tabular}}
\end{table*}
\rv{We recruited twelve participants (7 male, 5 female, average age of 28.3 years, SD=6.8) through contact with a disability organization.} Table \ref{table:Demographic background} shows the demographic information. We conducted interviews in person (n=2) and remotely (n=10) with participants who were not convenient offline.
To ensure participants can correctly understand the VR video content, we recruited participants with prior VR experience.

 Two participants with SMA-I exhibited restricted bodily mobility, limited exclusively to \rv{few fingers on each hand and} facial features such as eye, nose, and mouth movement \sout{as well as few fingers on each hand}. Three participants with SMA-II displayed unsteady hand movements and possessed partial control over their forearms, limited head, and trunk mobility, as well as constrained facial gestures. The remaining seven participants with SMA-III exhibited shaky hands and weakened hand and arm muscles, and they had difficulty lifting their hands above the chest level. Except for P8, all participants experienced difficulties in standing and walking. They relied on wheelchairs or \rv{bed} for daily mobility. Notably, all participants exhibited clear and fluent communication ability\sout{ and had prior experience with VR devices}. None of them had used upper-body gestures to control VR devices prior to the study. Participants received a compensation of \$15 for their participation.

\subsection{\rv{Commands Gathering}\sout{Tasks}}
To gather the VR commands and present the effects to participants,\sout{ we conducted a three-step analysis as shown in Figure \ref{fig:commands analysis}.}\rv{ we reviewed previous papers for VR commands but found none that included complete commands as a reference in our work. While there are three papers on user-defined VR gestures, each focuses on interactions within specific VR applications (e.g., CAD and open-world games \cite{moran-ledesmaUserDefinedGesturesPhysical2021}, VR shopping applications \cite{wuUserdefinedGestureInteraction2019}, and manipulating 3D objects \cite{nanjappanUserelicitedDualhandInteractions2018}) without addressing common VR interactions. }

\rv{Prior work of user-defined gestures in Section 2.3 (e.g., designed for smartphone \cite{zhaoDonWantPeople2022}, smartwatch \cite{arefinshimonExploringNontouchscreenGestures2016} and AR \cite{piumsomboonUserdefinedGesturesAugmented2013}) follows this two-step workflow to gather commands: 1) \textit{Gathering commands with high level of commonality across various applications}; 2) \textit{Joining the commands into several categories}. Inspired by this, we planned to follow a similar approach for gathering common VR commands. However, three key questions must be addressed before starting the process:  
\begin{itemize}
    \item Q1: How to select a diverse variety of VR applications?
    \item Q2: How to distill commonly used VR commands from various applications?
    \item Q3: How to objectively categorize VR commands?
\end{itemize}

}

\rv{
The three questions were answered through \textit{VR App Selection}, \textit{Video Analysis}, and \textit{Category Definition}, respectively. 
}

\begin{figure*}[ht]
    \centering
    \includegraphics[width=0.9\textwidth]{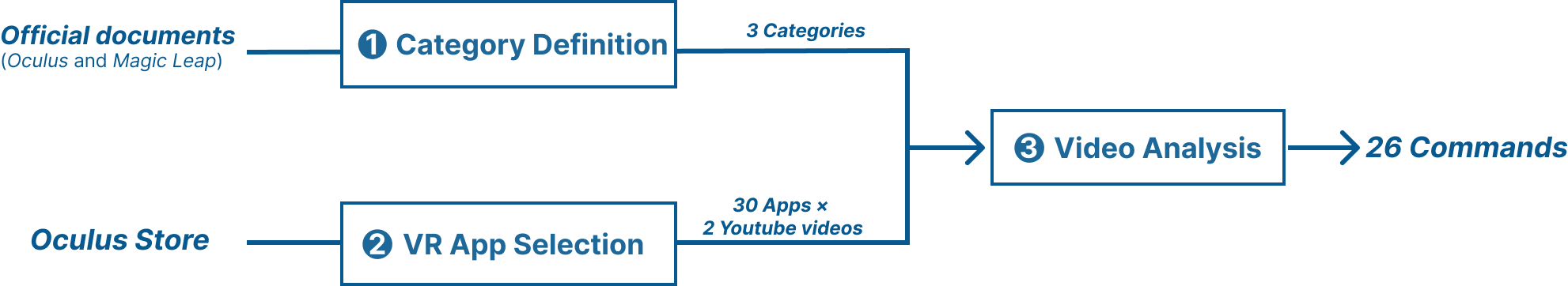}
    \caption{\textbf{The process for gathering the common VR commands: 1) \textit{Category Definition} established four categories of VR tasks,  named \textit{Selection}, \textit{Manipulation}, \textit{Menu-based Interaction} and \textit{Locomotion}; 2) \textit{Video Collection} specified VR tasks from popular VR application videos on YouTube; 3) \textit{Video Analysis} categorized the VR tasks into the four defined categories, following the established coding rules.}}
    \label{fig: commands analysis}
    \Description{Figure 2 demonstrates the three-step process for gathering the final set of VR commands:
1. Category Definition: The authors examined official VR documentation from Oculus, Magic Leap, and HTC VIVE to gain insights into categorizing VR tasks. They also drew inspiration from three research papers from Google Scholar. Through this, they distilled six fundamental categories of VR tasks: Pointing, Selection, Manipulation, Viewport Control, Menu-Based Interaction, and Locomotion.
2. Video Collection: The authors selected the top 10 most popular VR applications from each of the Games, Applications, and Entertainment categories on the Oculus Store homepage. This resulted in 30 VR applications. They searched each application name on YouTube and selected two 10+ minute videos from different publishers for each one, obtaining 60 videos total.
3. Video Analysis: Two researchers independently collected VR tasks by watching the videos and used an open coding method to group them. They established a coding rule that VR tasks with similar effects on the target object would share the same command code. Through iterative discussion and conflict resolution, they identified 26 common VR commands, categorized into menu-related, 3D object-related, and locomotion-related groups.}
\end{figure*}

\rv{In the \textit{Category Definition} phase, we consulted official documents (e.g., Oculus and Magic Leap) and previous research, ensuring objective references. However, in the other stages of \textit{VR App Selection} and \textit{Video Analysis}, all materials were chosen by the authors, introducing a potential subjective bias. To mitigate this bias, we restructured the workflow by prioritizing \textit{Category Definition}, where VR categories are established first, before proceeding to \textit{VR App Selection} and \textit{Video Analysis}.

In summary, as shown in Figure \ref{fig: commands analysis}, we conducted a three-step analysis to obtain the commands used in our study.}

1) \textbf{\textit{Category Definition}}. \sout{We closely examined official documents provided by Oculus\footnote{\url{https://developer.oculus.com/resources/hands-design-interactions/}} , Magic Leap\footnote{\url{https://ml1-developer.magicleap.com/en-us/learn/guides/design-interaction-overview}} , and HTC VIVE\footnote{\url{https://developer.vive.com/resources/vive-sense/hand-tracking-sdk/overview/}}  to gain insights into how to categorize VR tasks. Given the variability and lack of uniformity in standards across these websites, we also drew inspiration from three papers \cite{spittleReviewInteractionTechniques2023,bowmanTravelImmersiveVirtual1997,bowmanTestbedEvaluationVirtual1999} from Google Scholar. After then, we distilled six fundamental categories of VR tasks: Pointing, Selection, Manipulation, Viewport Control, Menu-Based Interaction and Locomotion. As shown in {\bfseries Table \ref{table:Fundamental Categories of VR Tasks}}.}\rv{To answer Q3, we referred to materials from both academia and industries to obtain VR command categories.} \rv{Initially, we looked into the recent work of Spittle et al \cite{spittleReviewInteractionTechniques2023}. This study conducted a systematic review of the state-of-the-art studies in immersive environments based on papers published between 2013 and 2020. It categorized seven immersive interaction tasks: pointing, selection, translation, rotation, scale, viewport, menu-based, and abstract. Then, we closely examined official documents from Oculus \footnote{\url{https://developer.oculus.com/resources/hands-design-interactions/}} and Magic Leap \footnote{\url{https://ml1-developer.magicleap.com/en-us/learn/guides/design-interaction-overview}} to further identify these interaction task categories from an industrial perspective. Specifically, we identified commonalities in the VR command categories mentioned in these materials and evaluated whether each category was suitable for VR gesture design to derive the final VR command categories. The complete recording of this process is presented in Appendix \ref{app:VR Categories}. Finally, we got four VR command categories shown in Table \ref{table: Fundamental Categories of VR Commands}.
}

\begin{table*}[ht]
    \caption{\textbf{Four Fundamental Categories of VR Commands}}
    \label{table: Fundamental Categories of VR Commands}
    \Description{Table 2. presents four fundamental categories of VR Commands: Selection, Manipulation, Menu-Based Interaction, and Locomotion. It provides a brief description for each category:
•Selection means initiating or confirming an action after pointing, such as grabbing an object up close or from a distance.
•Manipulation involves moving, rotating, or resizing interactive elements, as well as altering their properties.
•Menu-Based Interaction refers to presenting a structured set of tabs, commands, or utilities for users to engage with.
•Locomotion means moving or changing the direction of an avatar's position within a virtual space.}
    \resizebox{\textwidth}{!}{
    \begin{tabular}{l|lllllll}
    \hline
    \textbf{Categories of VR Commands} & \textbf{Description}\\ \hline
    Selection & Initiating or confirming an action after pointing, such as grabbing an object up close or from a distance.\\
    Manipulation & Moving, rotating, or resizing interactive elements, as well as altering their properties.\\
    Menu-Based Interaction & Presenting a structured set of tabs, commands, or utilities for users to   engage with.\\
    Locomotion & Moving or changing the direction of an avatar's position within a virtual space. \\
    \hline
    \end{tabular}}
\end{table*}

\sout{It is noted that \textit{Pointing} and \textit{Viewport Control} need to receive continuous real-time signals from the VR device, they are not well-suited for transforming into commands for user-defined gestures. Hence, we considered them as two interaction techniques and planed to gather participants' viewpoints on them in the interview section. Moverover, \textit{Selection} and \textit{Manipulation} often occur together while using VR, for example, users often grasp an object first and then manipulate it. Therefore, to assist participants to understand, we grouped \textit{Selection} and \textit{Manipulation} into one category named \textit{3D Object-related Commands}, and also named "Menu-Based Interaction" and \textit{Locomotion} as "Menu-related Commands" and "Locomotion-related Commands", respectively. }\rv{Furthermore, we made two modifications to the four VR categories candidates. Firstly, we incorporated \textit{distance}, a significant factor mentioned in the Oculus document \footnote{\url{https://developer.oculus.com/resources/hands-design-interactions/}}, as a subcategory to further divide each VR command category into \textit{near} and \textit{far}. This distinction is made because VR commands differ within or beyond the user's arm's reach (e.g., a near button can be selected by poking directly, but a far one can be selected by ray casting). Secondly, as \textit{Selection} and \textit{Manipulation} often occur together while using VR (e.g., users often grasp an object first and then manipulate it), we grouped them into one category named \textit{3D Object-related Commands}.
}
\begin{figure}[ht]
    \centering
    \includegraphics[width=0.6\linewidth]{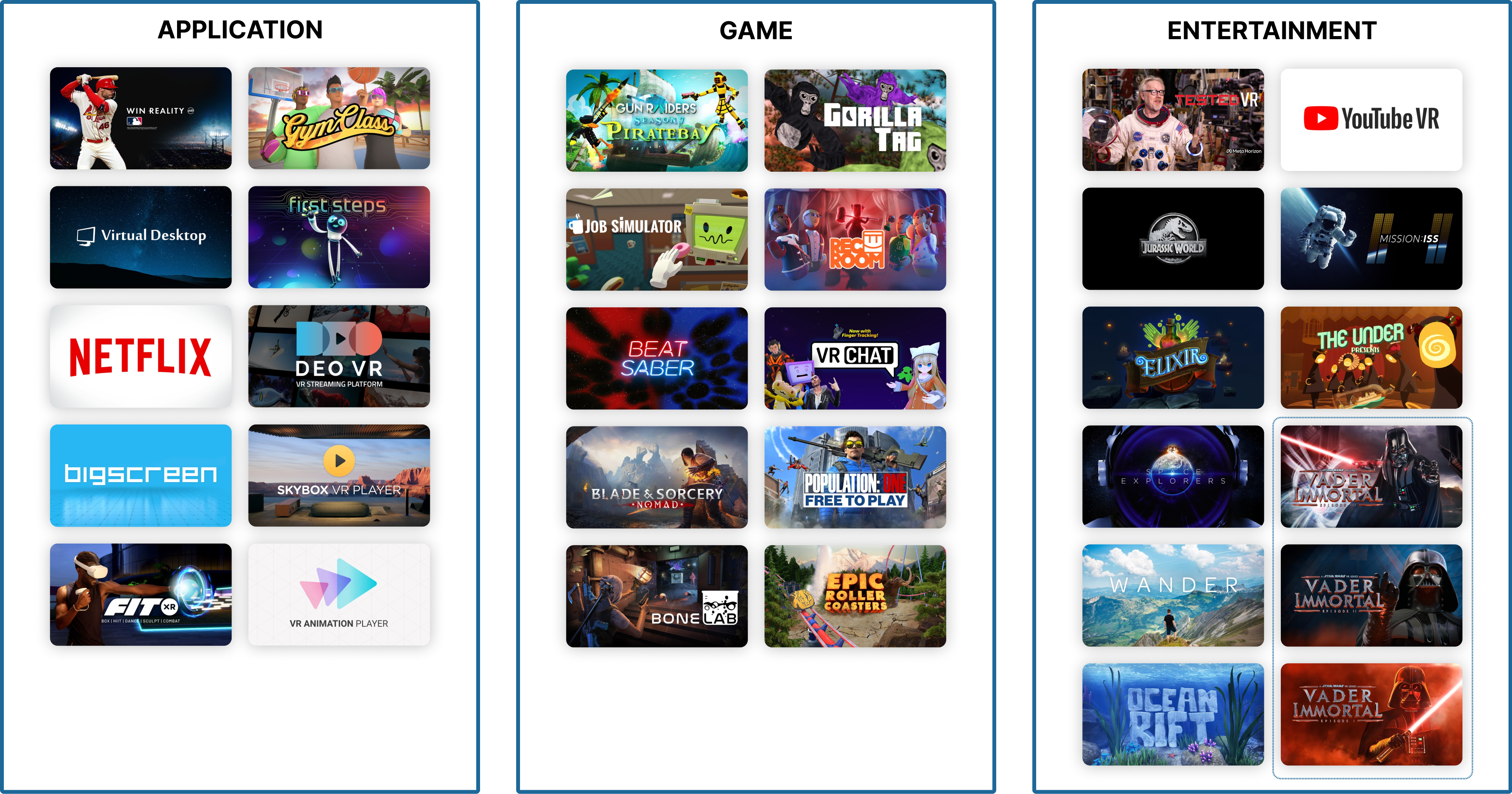}
    \caption{\textbf{30 VR applications in final. There are 32 VR apps in total, but in the Entertainment, \textit{Vader Immortal: Episode I, II, and III}  were regarded as one item because they had the same VR task but different contents.}}
    \label{fig:VR APPs}
    \Description{Figure 3 showcases a collection of 30 VR applications, which are organized into three categories: Application, Game, and Entertainment. The image displays a series of icons, each representing a different VR app or game.}
\end{figure}

    
    

2) \rv{\textbf{\textit{VR App Selection.}\sout{\textit{Video Collection}}} To address Q1, }we selected the Oculus Store as the app selection resource because its homepage categorizes VR applications into three tabs: Games \footnote{\url{https://www.oculus.com/experiences/quest/section/891919991406810/\#/?_k=bok987}}, Applications \footnote{\url{https://www.oculus.com/experiences/quest/section/1453026811734318/\#/?_k=8cv5gu}}, and Entertainment \footnote{\url{https://www.oculus.com/experiences/quest/section/841434313157491/\#/?_k=jxqawz}}\rv{, facilitating the choice of diverse applications. Each tab further contains several subtabs, such as Games for strategists, Kinetic sports, Music games, etc., with each subtab encompassing various VR applications. Focusing on the diversity of VR applications, we meticulously examined each subtab and identified the "Most Popular" subtab as the representative category for our study. We reasoned that "Popular," influenced by player preferences, is more likely to include diverse applications compared to other subtabs that concentrate on specific content (e.g., Games for strategists, which exclusively features applications related to strategies). }


We selected the top 10 most popular applications from \sout{each category}\rv{Games, Applications, and Entertainment} on August 13, 2023. \sout{In the Entertainment, \textit{Vader Immortal: Episode I, II, and III}  were regarded as one item because they had the same VR task but different contents. } Thus, we obtained $10\times3=30$ VR applications in final\rv{, as shown in Figure \ref{fig:VR APPs}}.\sout{ and searched for each application's name on YouTube, filtering the results by relevance. We selected two videos for each application, with a duration of over seven minutes, and from different publishers, resulting in a total of $30\times2=60$ videos}

3) \textbf{\textit{Video Analysis}}. \rv{
To answer Q2, we conducted a video analysis to collect common VR commands from various applications. 
}

\rv{To begin with, we searched for each application's name on YouTube, filtering the results by relevance. We selected two videos for each application, with a duration of over seven minutes, and from different publishers, resulting in a total of $30\times2=60$ videos. And then we collected VR tasks from those videos. Tasks refer to specific user actions in the context of each video. Based on these tasks (e.g., pulling a lever and pulling a drawer), commonly used VR commands will be distilled by summarizing and combining similar actions observed across various videos. }

\rv{
Secondly, we set a coding rule for better organizing VR tasks into several VR commands.} Two researchers independently collected VR tasks by watching the videos. However, we faced a challenge where a single command could correspond to different VR tasks depending on the context. \sout{For example, pulling a lever and pulling a drawer share similar motions but are treated as separate commands.}\rv{For instance, the action of pulling a lever upward and pulling a drawer backward, while distinct tasks share a similar motion involving flexing the forearm at the elbow joint. To address this ambiguity, }we drew inspiration from Vuletic et al.'s methodology in their systematic literature review of hand gesture types \cite{vuleticSystematicLiteratureReview2019}. \rv{They encountered similar ambiguity in coding hand gestures from different applied contexts while performing the same motion. Their rule for coding the gestures depends on their role or aim in the application.} We established a coding rule inspired by their work: \textit{VR tasks that have similar effects on the target object will share the same command code}. 

\rv{Finally,  }two researchers used an open coding method \cite{creswellQualitativeInquiryResearch2007} to group VR tasks and iteratively resolve conflicts through ongoing discussions. As a result, we identified 26 commands, as shown in Table \ref{table: The list of the commands}. 

\begin{table}[ht]
    \centering
    \footnotesize
    \caption{\textbf{The list of the 26 VR commands, grouped into three categories of Menu-related, 3D object-related and locomotion-related commands.}}
    \label{table: The list of the commands}
    \Description{Table 3 shows the list of the 26 VR commands which are grouped into three categories of Menu-related, 3D object-related, and locomotion-related commands. Each category contains the near interactions and the far interactions.}
    \renewcommand\arraystretch{1.35}
    \resizebox{0.48\textwidth}{!}{%
\begin{tabular}{c|c|c|c} 
\hline
\multicolumn{2}{c|}{\textbf{\textbf{Commands~}Category~}}                                                                                             & \textbf{No.} & \textbf{Commands}                                                                     \\ 
\hline
\multirow{8}{*}{\begin{tabular}[c]{@{}c@{}}Menu-related~\\ (\textit{Menu-based Interaction})\end{tabular}}                   & \multirow{2}{*}{Near}  & 1            & Access the Shortcut   Menu                                                            \\ 
\cline{3-4}
                                                                                                                             &                        & 2            & Confirm a Nearby   Selection                                                          \\ 
\cline{2-4}
                                                                                                                             & \multirow{6}{*}{Far}   & 3            & Access the Home Menu                                                                  \\ 
\cline{3-4}
                                                                                                                             &                        & 4            & Confirm a Far   Selection                                                             \\ 
\cline{3-4}
                                                                                                                             &                        & 5            & Scroll Up and Down                                                                    \\ 
\cline{3-4}
                                                                                                                             &                        & 6            & Scroll Left and Right                                                                 \\ 
\cline{3-4}
                                                                                                                             &                        & 7            & Zoom In or Zoom Out                                                                   \\ 
\cline{3-4}
                                                                                                                             &                        & 8            & Drag                                                                                  \\ 
\hline
\multirow{13}{*}{\begin{tabular}[c]{@{}c@{}}3D Object-related~\\ (\textit{Selection}~\& \textit{Manipulation})\end{tabular}} & \multirow{12}{*}{Near} & 9            & Grab a Nearby Object                                                                  \\ 
\cline{3-4}
                                                                                                                             &                        & 10           & Pinch a Nearby Object                                                                 \\ 
\cline{3-4}
                                                                                                                             &                        & 11           & Drop an Object                                                                        \\ 
\cline{3-4}
                                                                                                                             &                        & 12           & Throw an Object                                                                       \\ 
\cline{3-4}
                                                                                                                             &                        & 13           & Hit or Attack the   Target Object                                                     \\ 
\cline{3-4}
                                                                                                                             &                        & 14           & Chop the Target   Object                                                              \\ 
\cline{3-4}
                                                                                                                             &                        & 15           & Move the Target   Object                                                              \\ 
\cline{3-4}
                                                                                                                             &                        & 16           & Pull the Target   Object                                                              \\ 
\cline{3-4}
                                                                                                                             &                        & 17           & Rotate the Target   Object                                                            \\ 
\cline{3-4}
                                                                                                                             &                        & 18           & Shake or Swing the   Target Object                                                    \\ 
\cline{3-4}
                                                                                                                             &                        & 19           & Wave Towards the   Target Object                                                      \\ 
\cline{3-4}
                                                                                                                             &                        & 20           & \begin{tabular}[c]{@{}c@{}}Accumulate Force to \\~Hit the Target Object\end{tabular}  \\ 
\cline{2-4}
                                                                                                                             & Far                    & 21           & Grab a Distant Object                                                                 \\ 
\hline
\multirow{5}{*}{\begin{tabular}[c]{@{}c@{}}Locomotion-related\\ (\textit{Locomotion})\end{tabular}}                          & \multirow{4}{*}{Near}  & 22           & Artificial Locomotion                                                                 \\ 
\cline{3-4}
                                                                                                                             &                        & 23           & Jump                                                                                  \\ 
\cline{3-4}
                                                                                                                             &                        & 24           & Turn                                                                                  \\ 
\cline{3-4}
                                                                                                                             &                        & 25           & Lean or Bend                                                                          \\ 
\cline{2-4}
                                                                                                                             & Far                    & 26           & Teleportation                                                                         \\
\hline
\end{tabular}%
}
\end{table}
\subsection{Procedure}
\begin{figure*}[ht]
    \centering
    \includegraphics[width=0.95\linewidth]{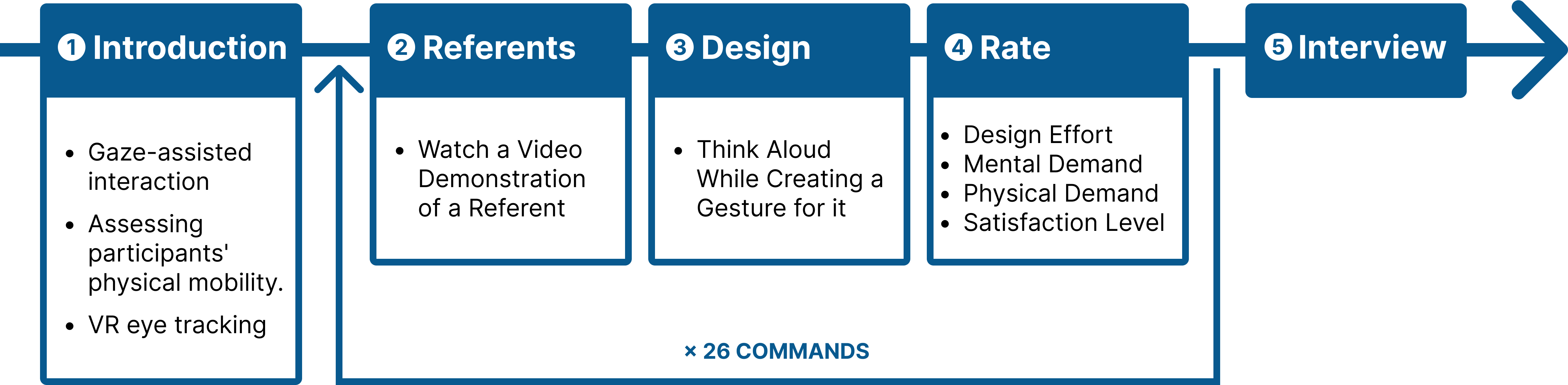}
    \caption{\textbf{The Study Procedure.}}
    \label{fig: study procedure}  
    \Description{Figure 3 illustrates the detailed procedure of the user study. The figure provides a detailed overview of the user-centered elicitation study procedure that engaged participants with SMA in designing gestures for VR tasks after presenting video stimuli as a reference. The key steps are:
Introduction Phase: Participants were first briefed on the project and asked to self-report their motor abilities. They were then presented with materials showcasing gaze-assisted interaction and viewport control in VR to confirm their comprehension.
Video Elicitation Phase: Participants watched brief video clips demonstrating the effect (referent) of each VR command one by one. The videos displayed synchronized actions in real life and first-person VR view.
Gesture Design Phase: After each video, participants designed and performed an upper-body gesture to accomplish the VR task. They used a think-aloud protocol during design.
Rating Phase: Participants rated the mental demand, physical demand, and performance of their gestures on a 7-point Likert scale.
Interview Phase: A semi-structured interview was conducted to understand participants' design rationales, preferences, and considerations regarding VR interaction.}
\end{figure*}

\rv {We used Tencent Meeting (a video conference application) to conduct and record all study sessions both in-person and remotely. All participants used personal computers, mobile phones, or tabletops to access the app and participate in our study. All participants, except P6, were positioned with one camera in front of them. The whole study procedure is shown in Figure \ref{fig: study procedure}}

\rv{1) \textbf{\textit{Introduction}}.} In the introduction phase, we first \rv {briefly introduced our project} and then asked participants to self-report their motor abilities, including their daily activities and challenges using various devices, to better understand their capabilities. \rv{During the participant's demonstration of their physical state, we observed the maximum and minimum range of their movements and guided the participant to either modify the camera's angle or adjust their distance from the camera. This ensured that all of the participant's gestures were comprehensively displayed in the video.} Then, participants were presented with materials including videos \footnote{Links of the videos:
\begin{itemize}
    \item \url{https://www.bilibili.com/video/BV1jk4y1p7ax/?share_source=copy_web&vd_source=f1455f0b9f86a68ed04b8784b03662bc}
    \item \url{https://youtu.be/NzLrZSF8aDM?si=79Q4iNuJ45JL5NBc}
    \item \url{https://youtu.be/5GTOU6e8--I?si=KL6v6tJRv0DLJPGv}
\end{itemize}
}, images, and text to illustrate the methods of gaze-assisted interaction in VR. After viewing these materials, we confirmed the participants' comprehension.

\begin{figure}[t]
     \centering
     \begin{subfigure}[b]{\textwidth}
         \centering
         \includegraphics[width=0.5\textwidth]{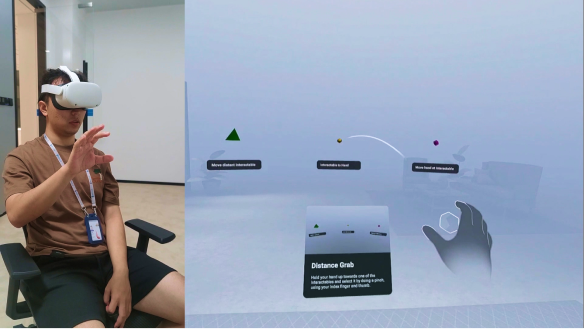}
         \caption{The video clip frame of \textit{Grab Distant Object}}
     \end{subfigure}
     \hfill
     \begin{subfigure}[b]{\textwidth}
         \centering
         \includegraphics[width=0.95\textwidth]{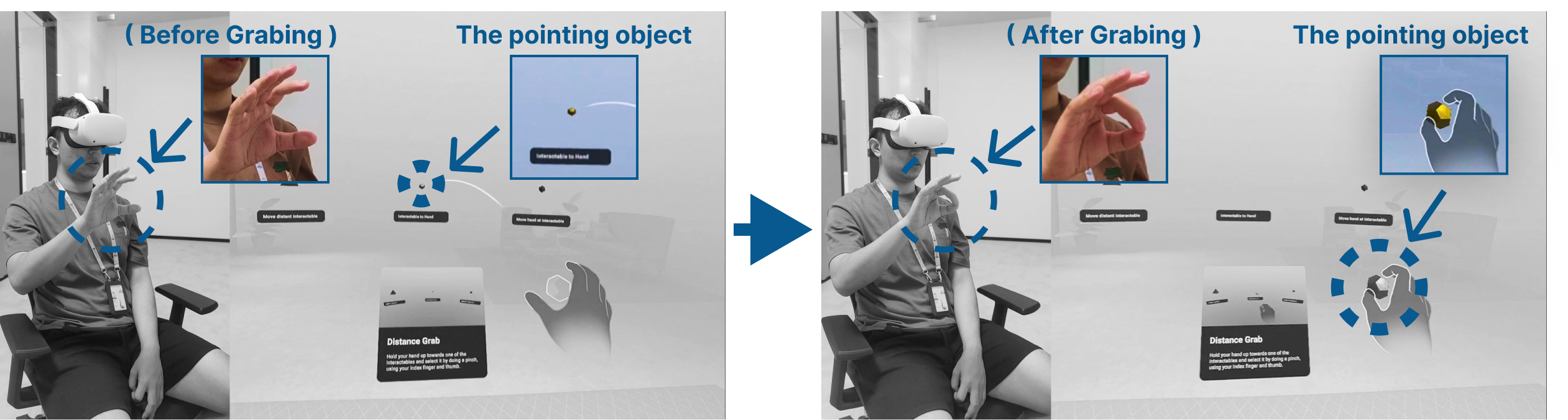}
         \caption{The video clip frame before and after \textit{Grab Distant Object}}
     \end{subfigure}
     \hfill
        \caption{\textbf{The effect of \textit{Grab Distant Object} command: (a) demonstrates the video clip frame of \textit{Grab Distant Object}. \rv{ One researcher wearing Quest2 recorded first-person views of grabbing a distant object in VR (on the right side of the screen), while another researcher captured his actions in reality (on the left side of the screen). These two videos were edited in time synchronization to assist participants in comprehending VR tasks}; (b) depicts the state before and after grabbing, showcasing an author's interaction with the VR system to execute a grabbing motion.}}
     \label{fig: video clip frame}
     \Description{Figure 5 shows an example video clip frame demonstrating the effect of the "Grab Distant Object" command. It contains:
(a) The video clip frame of Grab Distant Object, in which the actions of an author on the left are synchronized and consistent with the actions shown in the virtual environment on the right. This shows the user reaching out to grab an object in the distance.
(b) The before and after states of grabbing an object, depicting the user's first-person interaction with the VR system to execute a grabbing motion. On the left is the initial state with the hand empty. On the right is the grabbed state with the hand holding the target object.}
\end{figure}

\rv{2) \textbf{\textit{Referent Watching}}.} \rv{Participants were first exposed to brief video clips illustrating the effects of each VR command }(i.e., \textit{referent}, which is firstly called by Wobbrock et al. \cite{wobbrockUserdefinedGesturesSurface2009})\rv{, prior to designing gestures.} In Figure \ref{fig: video clip frame}, an example video clip frame for the "Grab Distant Object" command is shown \rv{(the video is available in the supplementary materials). Each video clip consisted of two parts, as depicted in Figure \ref{fig: video clip frame} (a). On the right side of the screen, a researcher wearing the Quest2 headset recorded first-person views of executing the command. On the left side, the researcher's movements while performing the VR command in reality were presented to aid understanding. These two perspectives were edited to be in time synchronization. It's important to note that the able-bodied movements shown in the video were not gestures but merely a component of VR command instruction. We emphasized this to participants, encouraging them to design gestures based on their own capabilities and preferences.}

\rv{3) \textbf{\textit{Gesture Design.}}} After each watching, participants were asked to create an upper-body gesture for the VR command and perform the gesture to the moderators. During this process, we asked participants to think aloud \rv{to verbalize their thoughts during the design process. Researchers inquired about participants' thought processes as shared during the design phase. For example, if a participant expressed the desire to create a cool gesture, moderators delved deeper, asking why they wanted to design gestures with cool physical appearances and what characteristics they considered cool. This approach aimed to gain a more profound and comprehensive understanding of the mental models of individuals with SMA in the VR gesture design process. }

 To reduce gesture conflicts, we asked participants to design different gestures for each command within the same \rv{category}\sout{(i.e., three \rv{categories} for Menu-based, 3D object-related, and locomotion-related commands)}. For commands that were in the different \rv{categories}, participants were allowed to perform the same gesture. \sout{However, due to the large number of commands, some participants might have forgotten their previous designs. To reduce gesture conflicts, we asked participants to design different gestures for each command within the same group (i.e., three groups for Menu-based, 3D object-related, and locomotion-related commands). For commands that were not in the same group, participants were allowed to design the same gesture.} However, due to a large number of commands, some participants might have forgotten their previous designs. Thus, a moderator monitored the gestures already created, and if she found a conflict in design, she would remind the participants to change either the current or previously designed gesture to a different one. Participants were allowed to change their previous gestures at any point during the process.

 \rv{4) \textbf{\textit{Rate the Gesture.}}} Upon completing each gesture design, participants were asked to rate four aspects of their gestures on a 7-point Likert scale, including design effort \rv{(the difficulty involved in designing gesture for the current referent)}, mental demand \rv{(the cognitive load required to execute the proposed gesture, such as memory)}, physical demand \rv{(the physical workload to execute the proposed gesture)}, and the overall satisfaction for the proposed gesture. \rv{Researchers followed up on these ratings to gain a deeper understanding of the participants' perspectives. For instance, if a gesture was rated high in physical demand but also high in overall satisfaction, researchers would inquire about the reasons behind these ratings.}

\rv{5) \textbf{\textit{Semi-structured Interview.}}} After the completion of all gesture designs, we conducted semi-structured interviews with each participant. These interviews were tailored based on the participant's responses during designing gestures, allowing us to ask follow-up questions for more in-depth information. The interview focused on the participants' experiences with accessibility issues in current VR device usage, their preferences, and main considerations when designing gestures, their expectations for VR interaction methods\rv{, and any additional content they wished to add. For example, some participants suggested the
desire for a universal gesture set integrated with eye movement and UI components, due to concerns about progressive
muscle atrophy in the future.}

\subsection{Data Analysis}
\rv{Our data analysis was divided into two parts. The first part involved categorizing and organizing user-defined gestures. The second part focused on a qualitative analysis of participants' mental models.}
\subsubsection{Gesture Analysis Method.} \rv{The original analysis methods for user-defined gestures \cite{wobbrockUserdefinedGesturesSurface2009} involved two main steps: \textit{First, classifying each gesture along four dimensions (form, nature, binding, and flow)} into a taxonomy to describe the gesture design space. \textit{Second, grouping identical gestures and selecting the group with the largest consensus as the representative gesture for each referent} for future design references.} \rv{Given our focus on SMA and VR, we made two significant changes, taking into account two factors: \textit{body parts and \textit{similar patterns.} Drawing inspiration from research on user-centered gesture design for individuals with motor impairments \cite{zhaoDonWantPeople2022}, which categorizes gesture taxonomy based on the involved body parts, we replaced \textit{four dimensions (form, nature, binding, and flow)} with \textit{the involved body parts}. We also referred to research on user-centered gesture design for VR \cite{wuUserdefinedGestureInteraction2019, moran-ledesmaUserDefinedGesturesPhysical2021}. This prompted us to shift the criteria from "gestures must be identical" to "gestures with a similar pattern," acknowledging the diversity in VR gesture design.}
\rv{The user-defined gestures were identified based on participants' descriptions and performed either through the camera or in person.}

\subsubsection{Mental Model Analysis Method.} The data consisted of audio recordings from online meetings, including participants' expressions during the think-aloud design phase, their explanations for scoring the design effort, mental demand, physical demand, and satisfaction level of their designed gestures, as well as their overall concerns and expectations expressed in the final semi-structured interview.} These recordings were transcribed into text scripts. Two researchers initially read through these scripts several times to gain an overall understanding of the participants' mental models in gesture design. Subsequently, the researchers independently coded the scripts using an open-coding approach \cite{creswellQualitativeInquiryResearch2007}. \rv{Themes, subthemes, and specific contents were inductively constructed by assigning keywords to participants' responses.} Repeating or similar keywords were grouped into higher-level categories. For example, the sub-theme "Concerns about the Accuracy of Recognition of Micro Gestures" emerged when phrases like "wrong recognition," "worried about recognition," and "unrecognition" frequently appeared. The coders regularly discussed and reconciled any coding discrepancies. Further meetings with other co-authors were conducted to finalize agreements based on the preliminary coding. Ultimately, we identified four mental models of participants with SMA, which are detailed in Section 5: Mental Model Observations. \rv{A codebook of mental models is in the supplementary materials.}


%% file: Results.tex
\rv{ Our results were divided into two parts. The first part shows the classification results of user-defined gestures. The second shows four mental models of people with SMA when designing VR gestures.}

\subsection{Results of Gesture Analysis}
\begin{figure*}[ht]
    \centering
    \includegraphics[width=\linewidth]{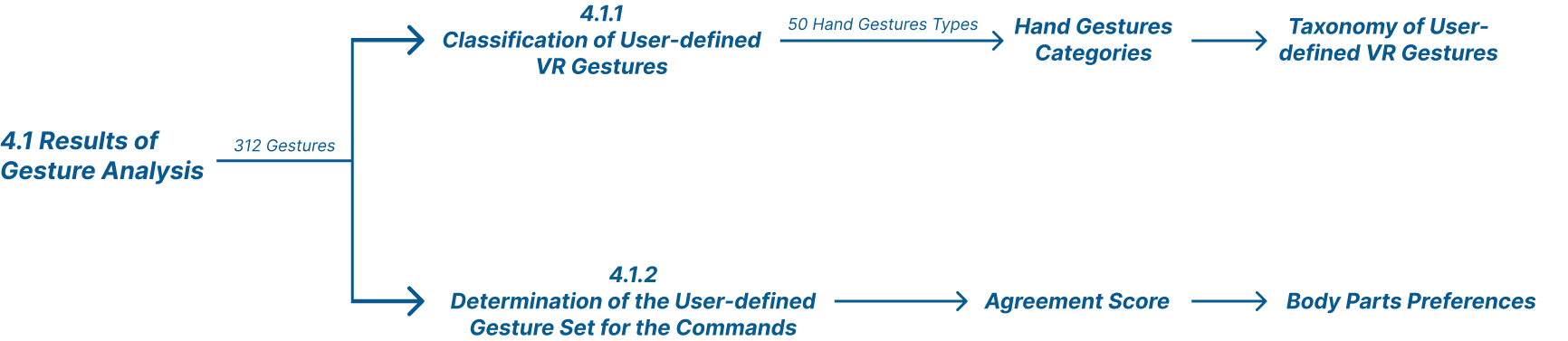}
    \caption{\textbf{The Processing of User-defined Gesture Analysis}}
    \label{fig: Results Analysis Methods}  
    \Description{Figure 6 is a flowchart titled "The Processing of User-defined Gesture Analysis," which outlines the analysis process in sequential steps:4.1 Results of Gesture Analysis:This initial box references the analysis of 312 gestures.4.1.1 Classification of User-defined VR Gestures:The process starts with the classification of user-defined VR gestures, which led to the identification of 50 hand gesture types.Hand Gestures Categories:Following the classification, these gestures are further categorized, the specifics of which are not detailed in the description.
Taxonomy of User-defined VR Gestures:The categorized gestures are then used to develop a taxonomy of user-defined VR gestures.4.1.2 Determination of the User-defined Gesture Set for the Commands:The next step involves determining the user-defined gesture set for the commands.
Agreement Score:An agreement score is calculated, likely indicating the level of consensus or standardization of the gestures among users.
Body Parts Preferences:
The final step in the flowchart focuses on identifying body parts preferences, possibly indicating which parts of the body users prefer to use when making gestures.}
\end{figure*}

 Due to the diverse personalized preferences of user-defined gestures, we were unable to finalize a specific set of gestures. However, we developed a taxonomy and gained insights into participants' usage of different body parts by analyzing the distribution of gestures. The structure of this section is depicted in Figure \ref{fig: Results Analysis Methods}.
 
\subsubsection{Classification of User-defined VR Gestures} \rv{We collected a total of 312 gestures (12 Participants × 26 Commands).} 
\begin{figure}[H]
    \centering
    \begin{subfigure}[t]{0.29\textwidth}
        \centering
        \includegraphics[width=\textwidth]{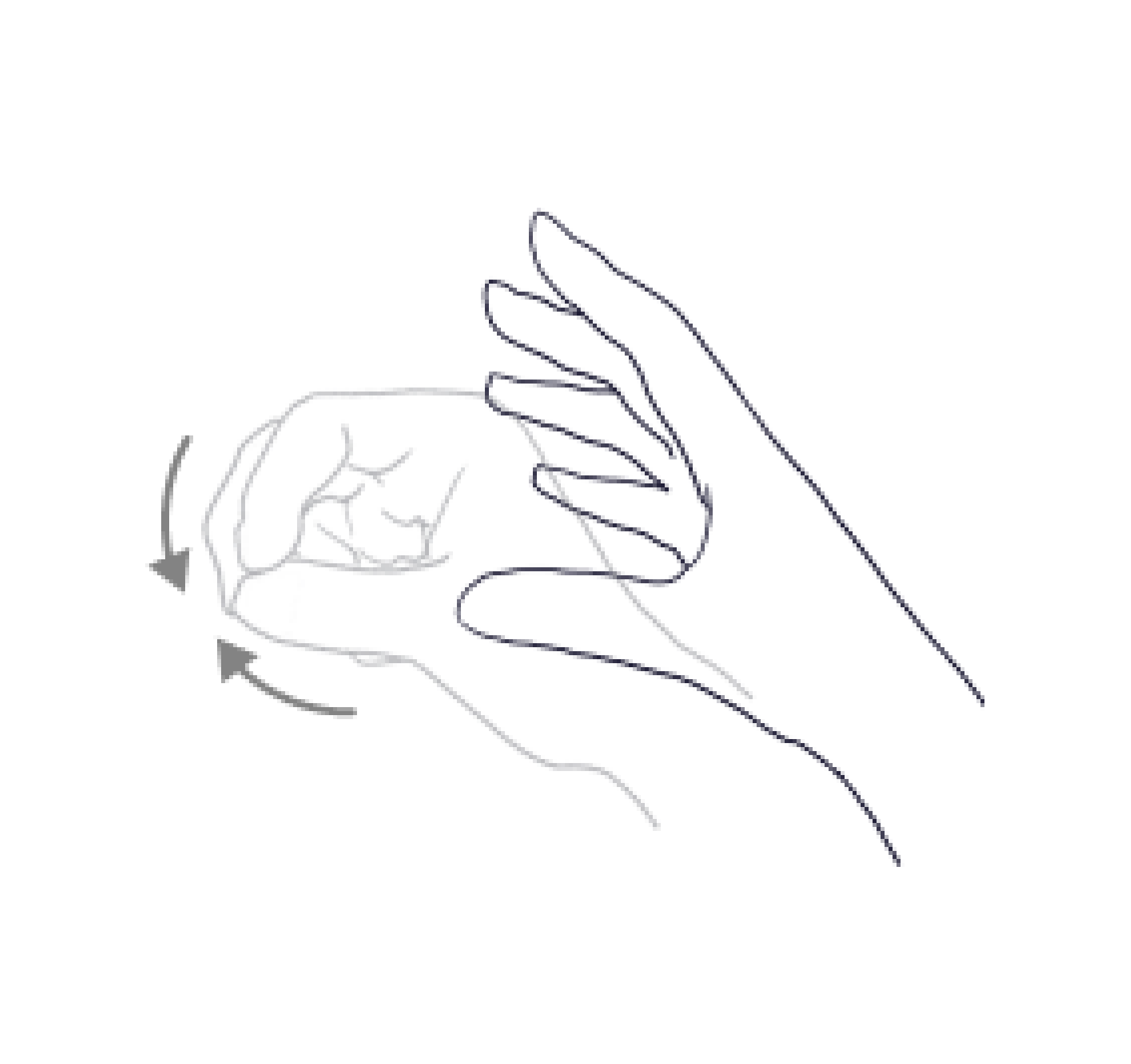}
        \caption{Thumb-to-finger Gesture type.}
    \end{subfigure}
    \begin{subfigure}[t]{0.29\textwidth}
        \centering       \includegraphics[width=\textwidth]{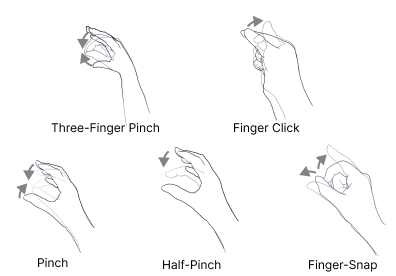}
        \caption{Five specific user-defined gestures of Thumb-to-finger Gesture type.}
    \end{subfigure}
    \caption{\textbf{Example of coding Thumb-to-finger Gesture Type: (a) illustrates the representative motion pattern of the Thumb-to-finger Gesture type, where the thumb touching the fingers indicates confirmation or precise control. (b) shows five user-defined gestures. Furthermore, these five gestures are grouped as the Thumb-to-finger Gesture, as shown in (a).}}
    \label{fig:Gesture Coding} 
    \Description{Figure 7 shows the examples of coding Thumb-to-finger Gesture Type. Figure5 (a) illustrates the representative motion pattern of the Thumb- to-Finger Gesture type, where the thumb touching the fingers indicates confirmation or precise control. Figure 5 (b) shows five user-defined gestures, including Three-Finger Pinch, Finger Click, Pinch, Half-Pinch, and Finger-Snap. These five gestures are grouped as the Thumb-to-Finger Gesture, as shown in (a)}
\end{figure}

    \textbf{Hand Gestures Categories.} \rv{  While constructing the taxonomy based on body parts and movement patterns, we found a vast diversity in hand gestures. Our collection of 126 hand gestures, making up 40.3\% of the total 312 gestures, included 50 different types. To categorize them into taxonomy breakdowns, we further refined the diverse hand gestures based on specific hand parts (e.g., thumb, index finger, palm, etc.) and similar movement patterns (e.g., swipe, slide, grip, strike, etc.).} Figure \ref{fig:Gesture Coding} illustrates an example of this process. We grouped five gestures, including the three-finger pinch, thumb-to-index pinch, finger click, half-pinch, and finger snap, into the thumb-to-finger gesture type.

    \rv{Finally, we identified five basic hand gesture patterns, as shown in Table \ref{tab: Hand Gesture Identification},} including thumb-to-finger gestures(TFG), grasping and closure gestures (GCG), swinging and extension gestures (SEG), tapping and striking gestures (TSG), shaping and symbolizing gestures (SSG). \rv{The complete visual illustration can be seen in Appendix \ref{app:The complete visual illustration of Hand Gestures Categories} Figure  \ref{fig:The complete visual illustration of Hand Gestures Categories.}.}
\begin{table*}[h]
    \renewcommand\arraystretch{1.15}
    \caption{\textbf{The identification of five hand gesture types.}}
    \label{tab: Hand Gesture Identification}
    \Description{ategorizes hand gestures into five distinct types, each with a description and examples for better understanding. Thumb-to-Finger Gesture:
1. Hand gestures involving the thumb touching the fingers signify confirmation or precise control.Examples: Thumb-to-index Pinching, Three-Finger Pinch, Thumb-to-index Clicking, Thumb-to-index Half-pinching, Thumb-to-index Snapping quickly.
Grasping or Closure Gestures:
2. Hand gestures that involve the proximity or closure of the palm or fingers convey a grasp or closing motion.Examples: Five-Finger Grasping, Clenching Fist, Four-Finger Picking, index-finger hooking.
Swinging or Extension Gestures:
3. Hand gestures that involve the spreading or flicking of the palm or fingers indicate a direction or signal a return to the initial state.Examples: Index finger sliding to the left or right, Thumb sliding to the left or right, Hand palm opening, Index and thumb spreading apart, Hand palm swinging.
Tapping or Striking Gestures:
4.Hand gestures that involve tapping or lightly tapping another surface in the air with the surface of the hand convey a clicking or tapping motion. Examples: Both hands clapping together, Index finger tapping or pressing, One hand’s fist striking the other hand.
Symbolic Hand Signs:
5. Hand gestures that utilize specific hand or finger positions convey specific meanings or information. Examples: Forming the "OK" sign with fingers, Single finger moving in a circular motion, Fingers forming the shape of the number "6", Fingers simulating a gun shape, Fingers outlining the shape of a person.
Each type of gesture serves a different communicative purpose and is illustrated with examples to showcase the diversity of hand gestures used in interactions.}
    \resizebox{1\textwidth}{!}{%
        \begin{tabular}{l|l|l}
        \hline
        \textbf{Hand Gestures Type}    & \textbf{Description}                                                                                                                                                                           & \textbf{Example}                                                                                                                                                                                                                                             \\ \hline
        Thumb-to-Finger Gesture        & \begin{tabular}[c]{@{}l@{}}Hand Gestures that involve the thumb touching \\ the fingers signify confirmation or precise control.\end{tabular}                                                  & \begin{tabular}[c]{@{}l@{}}Thumb-to-index Pinching, Three-Finger Pinch,\\ Thumb-to-index Clicking, \\ Thumb-to-index Half-pinching, \\ Thumb-to-index Snaping quickly, etc.\end{tabular}                                                                     \\ \hline
        Grasping or Closure Gestures   & \begin{tabular}[c]{@{}l@{}}Hand Gestures that involve the proximity or closure\\ of the palm or fingers convey a grasp or closing motion.\end{tabular}                                         & \begin{tabular}[c]{@{}l@{}}Five-finger Grasping, Clenching Fist, \\ Four-finger Picking, index-finger hooking, etc.\end{tabular}                                                                                                                             \\ \hline
        Swinging or Extension Gestures & \begin{tabular}[c]{@{}l@{}}Hand Gestures that involve the spreading or flicking\\ of the palm or fingers indicate a direction or signal a\\ return to the initial state.\end{tabular}          & \begin{tabular}[c]{@{}l@{}}Index finger sliding to the left or right, \\ Thumb sliding to the left or right,\\ Hand palm opening, Index and thumb spreading apart, \\ Hand palm swinging, etc.\end{tabular}                                                  \\ \hline
        Tapping or Striking Gestures   & \begin{tabular}[c]{@{}l@{}}Hand Gestures that involve tapping or lightly tapping\\ another surface in the air with the surface of the hand\\ convey a clicking or tapping motion.\end{tabular} & \begin{tabular}[c]{@{}l@{}}Both hands clapping together, \\ Index finger tapping or pressing, \\ One hand's fist striking the other hand, etc.\end{tabular}                                                                                                  \\ \hline
        Symbolic Hand Signs            & {\color[HTML]{111111} \begin{tabular}[c]{@{}l@{}}Hand Gestures that utilize specific hand or finger \\ positions convey specific meanings or information.\end{tabular}}                        & \begin{tabular}[c]{@{}l@{}}Forming the "OK" sign with fingers, \\ Single finger moving in a circular motion, \\ Fingers forming the shape of the number "6", \\ Fingers simulating a gun shape, \\ Fingers outlining the shape of a person,etc.\end{tabular} \\ \hline
        \end{tabular}
    }
    \end{table*}

\textbf{Taxonomy of User-defined VR Gestures.}  We identified a user-defined gestures taxonomy, as shown in Table \ref{tab: Taxonomy}. We grouped the gestures into 15 categories based on the body parts involved. These 15 categories included \textit{a single body part} and \textit{the combinations of different body parts}: only hand, \sout{only} wrist, \sout{only} forearm, only shoulder, \sout{only} torso \rv{or chest}, only head, only eyes, only mouth, hand \&\& wrist, hand \&\& forearm, head \&\& mouth, eyes \&\& hand, eyes \&\& wrist, eyes \&\& mouth and others. \sout{Among all categories, the Only Hand and Only Eyes category were the most diverse. }
    
    The \textit{Only Hand} category includes\sout{ five basic types of hand gestures:} \textit{Thumb-to-Finger Gestures (TFG), Grasping \rv{or}\sout{and} Closure Gestures (GCG), Swinging \rv{or}\sout{and} Extension Gestures (SEG), Tapping \rv{or}\sout{and} Striking Gestures (TSG), Shaping \rv{or}\sout{and} Symbolizing Gestures (SSG)} and the combinations of \rv{them}\sout{these basic hand gestures}. \rv{The combination of two basic hand gestures refers to the execution of two hand movements either subsequently or simultaneously. For instance, the \textit{thumb-to-index pinch and horizontal slide} can be a sequential combination in TFG+SEG, where the thumb-to-index pinch belongs to TFG and the horizontal slide belongs to SEG.} \sout{TFG involves the thumb touching the fingers and signifies confirmation or precise control. GCG conveys a grasp or closing motion through the proximity or closure of the palm or fingers. SEG shows a direction or signals a return to the start by spreading or flicking the palm or fingers. TSG involves lightly tapping another surface with the hand. SSG utilizes specific hand or finger positions to convey specific meanings or information.}

\begin{table*}
\renewcommand\arraystretch{1.12}
\caption{\textbf{The fifteen categories of the user-defined upper-body gestures and the gestures types within each category.}}
\label{tab: Taxonomy}
\Description{Table 5 shows the fifteen categories of the user-defined upper-body gestures and the gestures type within each category. These gestures are grouped into 15 categories based on the body parts involved. These 15 categories included a single body part and the combinations of different body parts: only hand, only wrist, only forearm, only shoulder, only torso, only eyes, only mouth, only head, hand && wrist, hand && forearm, hand && mouth, eyes && hand, eyes && wrist, eyes && mouth and others.}
\resizebox{0.9\textwidth}{!}{%
    \begin{tabular}{c|c|c|c}
    \hline
    \textbf{Taxonomy}               & \textbf{Breakdown}                     & \textbf{Taxonomy}                 & \textbf{Breakdown}                       \\ \hline
                                    & Thumb-to-Finger Gesture (TFG)          &                                   & Eye Movement                             \\ \cline{2-2} \cline{4-4} 
                                    & Grasping \rv{or}\sout{and} Closure Gestures (GCG)    &                                   & Wink                                     \\ \cline{2-2} \cline{4-4} 
                                    & Swinging \rv{or}\sout{and} Extension Gestures (SEG)  &                                   & Blink                                    \\ \cline{2-2} \cline{4-4} 
                                    & Tapping \rv{or}\sout{and} Striking Gestures (TSG)    &                                   & Eye Size                                 \\ \cline{2-2} \cline{4-4} 
                                    & Shaping \rv{or}\sout{and} Symbolizing Gestures (SSG) &                                   & Gaze Dwell                               \\ \cline{2-2} \cline{4-4} 
                                    & GCG/SEG + TSG                          &                                   & Eyebrows                                 \\ \cline{2-2} \cline{4-4} 
    \multirow{-7}{*}{Only Hand}     & GCG/TFG + SEG                          &                                   & Eyebrows + Eye Movement                  \\ \cline{1-2} \cline{4-4} 
                                    & Wrist Hook and Swing                   & \multirow{-8}{*}{Only Eyes}       & Wink/ Eye Movement/Gaze Dwell + Eye Size \\ \cline{2-4} 
    \multirow{-2}{*}{\sout{Only }Wrist}    & Wrist Rotate                           &                                   & Lip \rv{Movement}\sout{Gesture}                              \\ \cline{1-2} \cline{4-4} 
                                    & Forearm Swipe                          & \multirow{-2}{*}{Only Mouth}      & Mouth \rv{Open then Close}\sout{Size}                               \\ \cline{2-4} 
    \multirow{-2}{*}{\sout{Only }Forearm}  & Forearm Swing                          &                                   & SEG/TSG/SSG + Wrist Hook and Swing       \\ \cline{1-2} \cline{4-4} 
                                    & Shoulder Shrug                         & \multirow{-2}{*}{Hand \& Wrist}   & TFG/GCG/SEG/SSG + Wrist Rotate           \\ \cline{2-4} 
    \multirow{-2}{*}{Only Shoulder} & Shoulder Alternation Shake             &                                   & TFG/GCG/SEG/TSG/SSG + Forearm Swing      \\ \cline{1-2} \cline{4-4} 
                                    & Torso Rotation and Tilt                & \multirow{-2}{*}{Hand \& Forearm} & GCG/SEG/TSG/SSG + Forearm Swipe          \\ \cline{2-4} 
    \multirow{-2}{*}{\sout{Only }Torso \rv{or Chest}}    & Chest \rv{Lifting}\sout{Protrusion}                       & Head \& Mouth                     & Head Tilt + Lip \rv{Movement}\sout{Gesture}                  \\ \hline
                                    & Head Turn                              & Eyes \& Hand                      & GCG/SEG + Eye Movement                     \\ \cline{2-4} 
                                    & Head Nod                               & Eyes \& Wrist                     & Eye Size + Wrist Hook and Swing          \\ \cline{2-4} 
                                    & Head Tilt                              & Eyes \& Mouth                   & Eyebrows + Mouth \rv{Open then Close}\sout{Size}                    \\ \cline{2-4} 
    \multirow{-4}{*}{Only Head}     & Head Turn + Nod                        & Others                            & Eyes/Hand/Mouth + UI                     \\ \hline
    
    \end{tabular}}
\end{table*}

    \rv{The \textit{Wrist} category includes two basic types: \textit{Wrist Hook and Swing} and \textit{Wrist Rotate}. \textit{Wrist Hook and Swing} refers to movements of the wrist around the wrist joint in various directions. Additionally, \textit{Wrist Rotate} typically includes the twisting of the hand around its axis. The \textit{Forearm} category includes \textit{Forearm Swipe} and \textit{Forearm Swing}, with variations in the extent of forearm rotation centered around the elbow joint. }

    \rv{The \textit{Only Shoulder} category includes \textit{Shoulder Shrug} and \textit{Shoulder Alternation Shake}. \textit{Shoulder Alternation Shake} indicates the sequential and alternating forward movement of each shoulder, generating a shaking motion. The \textit{Torso or Chest} category includes \textit{Torso Rotation and Tilt} and \textit{Chest Lifting}. }

    \rv{The \textit{Only Head} category includes \textit{Head Turn, Head Nod, Head Tilt}, as well as \textit{Head Turn and Nod}.}

    The \textit{Only Eyes} category includes eight basic types of eye gestures: \textit{Eye Movement}, \textit{Wink}, \textit{Blink}, \textit{Eye Size}, \textit{Gaze Dwell}, \textit{Eyebrows}, and the combinations of these basic eye gestures. \textit{Eye Movement} includes moving the eyes up and down, left and right, or eye rotating. \rv{\textit{Wink} and \textit{Blink} respectively indicate the rapid closure and reopening of one eye or both eyes.}\sout{\textit{Wink} is a brief, intentional closure of one eye. \textit{Blinks} include single and double blinks, as well as different numbers of blinks.} \textit{Eye Size} includes wide opening, closing, and squinting of the eyes. \textit{Gaze Dwell} is the act of maintaining prolonged eye contact with the target object. \sout{\textit{Eyebrow} includes squeezing eyebrows. }\rv{\textit{Eyebrow} includes movements associated with raising or lowering the eyebrows.}

     \rv{The \textit{Only Mouth} category includes \textit{Lip Movement} and\textit{ Mouth Open then Close}. \textit{Lip Movement} refers to various actions or changes primarily occurring in the area of the lips, such as \textit{Wry Mouth} and \textit{Pucker Lips}.}

     \rv{The \textit{Others} category, introduced by P1 with SMA-I, diverges from the approach of designing unique gestures for each command. P1 advocates placing commands on menus or icon-based buttons and using available movements for selection.}


   

    \rv{\subsubsection{Determination of the User-defined Gesture Set for the Commands}
To derive the final user-defined gesture set from all gestures proposed by all participants, we first collated the gestures included in each command and counted the number of participants performing the same gesture. The number was also used to calculate the agreement score of the commands. }

 \rv{\textbf{Agreement Score.}} The agreement score was initially proposed by Wobbrock et al. \cite{wobbrockUserdefinedGesturesSurface2009} and later widely used in studies uncovering user-defined gestures. It intuitively characterizes differences in agreement between target users for assigning a gesture to a given command. In general, the higher the agreement score of a command, the better the participants are in agreement with the gesture assigned to the command. \rv{We used the following equation to calculate the agreement score from prior user-defined gesture research \cite{ruizUserdefinedMotionGestures2011,wobbrockUserdefinedGesturesSurface2009}:}

\begin{equation}
     A_{c} =\sum_{ P_{i}}( \frac{P_{i}}{P_{c}})^2    
\end{equation}
    In Equation 1, $c$ is one of the commands, $A_c$ represents its agreement score based on participants’ proposed gestures for this command. $P_c$ is the total number of gestures proposed for c, which is the number of participants in our case (N=12). $i$ represents a unique gesture. $P_i$ represents the number of participants who propose the unique gesture $i$. Take the \textit{Confirm a Far Selection} command as an example, 12 participants proposed 12 gestures in total, $P_c$ equals 12. Among these gestures, there were 9 unique gestures: 3 (Blink \rv{Twice}), 2 (\rv{Head Nod Once}), 1 (\rv{Gaze for 5-10s}), 1 (\rv{Furrowed Brow} + Wink \rv{the Right Eye}), 1 (\rv{Pout}), 1 (\rv{Fingers simulating a Gun Shape}), 1 (\rv{Pinch}), 1 (\rv{Index Finger Tapping}), and  1 (\rv{Thumb Swing}). As a result, the agreement score of the \textit{Select Far Selection Button} command was calculated as follows:
    \begin{equation}
        \left ( \frac{3}{12}\right )^{2}+\left ( \frac{2}{12}\right )^{2}+7\left ( \frac{1}{12}\right )^{2}= 0.14 
    \end{equation}

\begin{figure}[!h]
    \centering
    \includegraphics[width=0.5\linewidth]{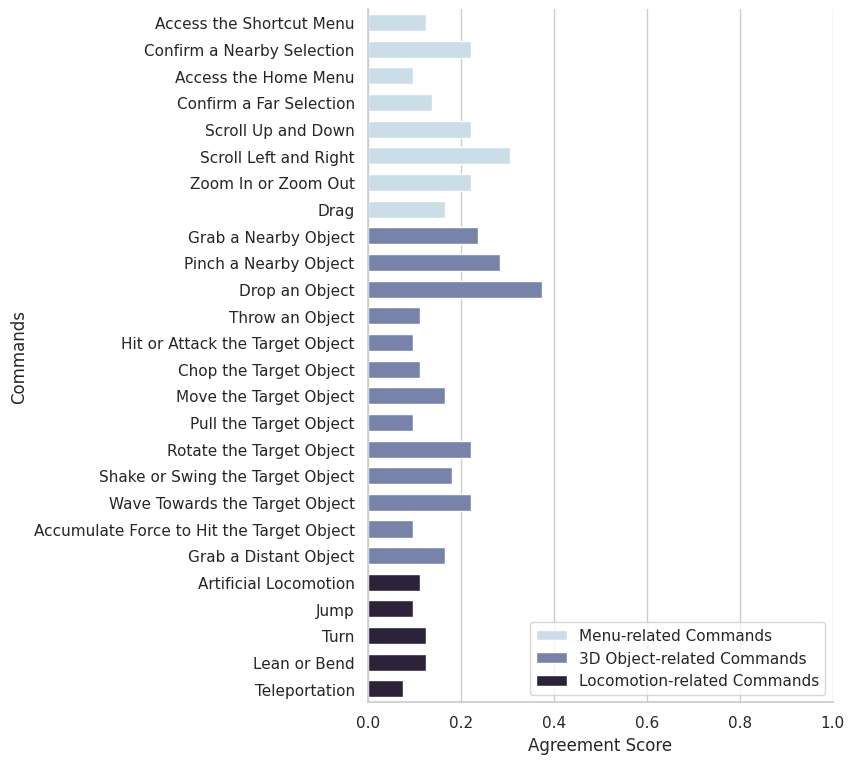}
    \caption{\textbf{The agreement scores of the 26 commands. The higher the score, the higher the participants' consensus on which gesture type should be assigned.}}
    \label{fig: AgreementScore}    
    \Description{Figure 6 shows the agreement scores of the 26 commands which are categorized into Menu-related Commands, 3D Object-related Commands, and Locomotion-related Commands. The agreement score is ranged from 0 to 1. The higher the score, the higher the participants’ consensus on which gesture type should be assigned.}
\end{figure}

\begin{figure*}[ht]
    \centering
    \includegraphics[width=\linewidth]{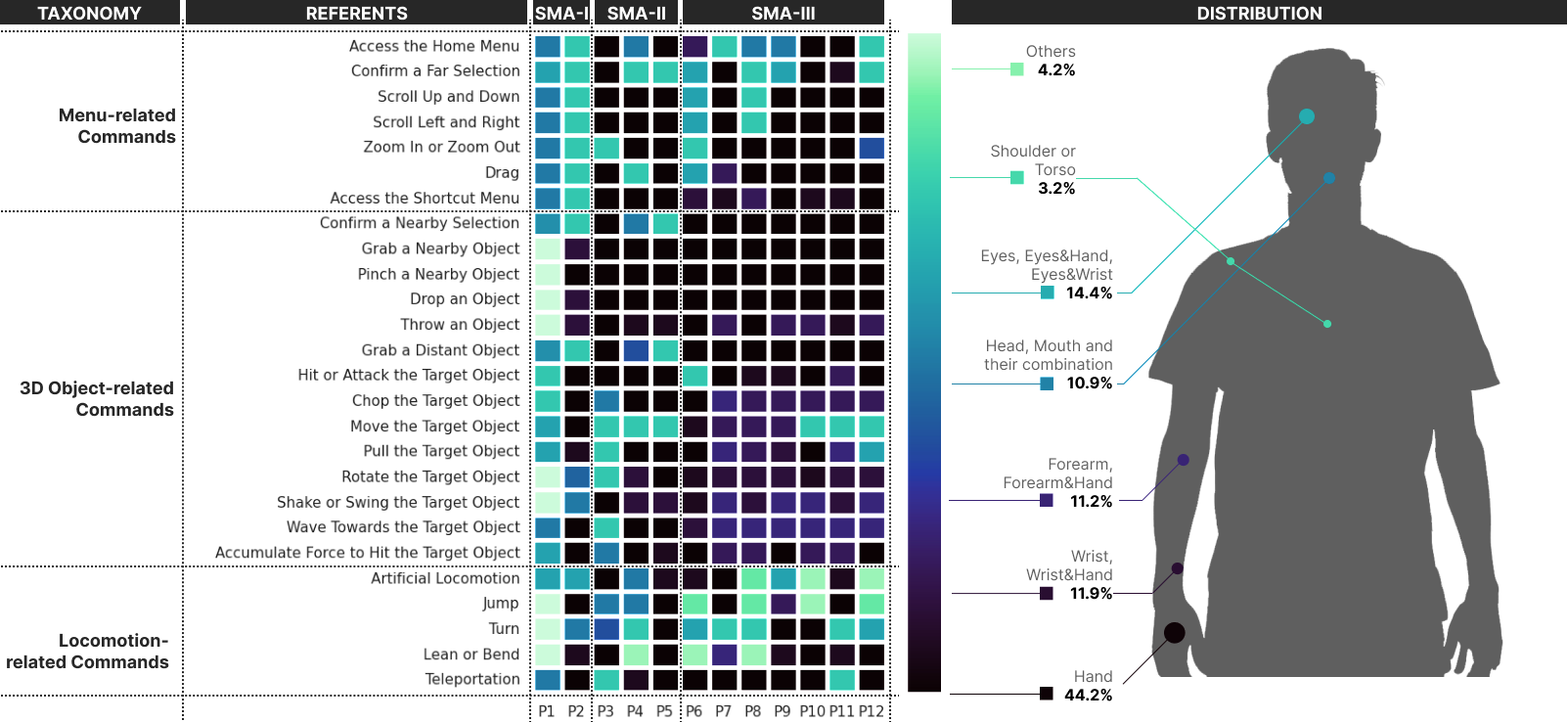}
    \caption{\textbf{Participant-Designed Gesture Categories for VR Commands and the Distribution of Gesture Groups.}}
    \label{fig:Distribution}  
    \Description{Figure 9 shows the Participant-Designed Gesture Categories for VR Commands and the Distribution of Gesture Groups. Among these gestures, 44.2\% were only-hand gestures, 12.8\% were only-eyes gestures, 5.8\% were only-wrist gestures, 4.5\% were only-forearm gestures, 4.2\% were only-head gestures, 6.1\% were only-mouth gestures, 1.3\% were only-arm gestures, 1.9\% were only-torso gestures, 6.1\% were hand & wrist, 6.7\% were hand & forearm, 1.3\% were eyes & hand, 2.9\% were mouth & UI and the rates of the other combined gestures were less than 1\%.}
\end{figure*}

   Figure \ref{fig: AgreementScore} shows the agreement score of the gestures proposed for each command.  \rv{For most commands, the agreement score is low, which indicates that the participants proposed diverse gestures for most commands and less agreed on which gesture should be allocated to them. The agreement score of \textit{Teleportation} is lowest since every participant proposed different gestures in this command and this indicates no common gesture can be allocated to it. The diversity of gestures makes obtaining a common user-defined gesture set challenging. To better understand participants’ preferences when designing VR gestures, we categorized the body parts used by each participant for each command, as shown in Figure \ref{fig:Distribution}.}

   \rv{\textbf{Body Parts Preferences.} Figure \ref{fig:Distribution} indicates that although the overall trend is that individuals with higher motor abilities use upper limb body parts more, there are differences observed among different participants and VR commands. For example, even participants with SMA-I, P1, and P2 preferred to use different body parts (refer to the first 2 columns in Figure \ref{fig:Distribution}).} P1\sout{had almost completely lost his hand mobility and} relied more on body parts above the neck\sout{, especially the eyes,} to design gestures. In cases where \rv{above-the-neck} body parts were insufficient or appropriate \sout{for tasks}, P1 used UI combined with \rv{body parts} to solve the {\itshape "shortage of body parts to design"} (P1). Despite also having limited hand mobility, P2 still wanted to use his hands to design gestures for a better sense of body involvement \rv{in VR}. He chose to use peripherals such as a mouse (which he also used in daily life for computer operations, as shown in Figure \ref{fig:teaser} P2) to support his hand muscles and perform small hand movements \rv{to design gestures}.

The body parts used also varied for different categories of tasks. \rv{For \textit{Menu-related Commands}, the proportion of using the eyes is higher compared to other categories. For \textit{3D Object-related Commands}, the hands are being used more among all participants except P1. For \textit{Locomotion-related Commands}, the body parts chosen were more diverse.}

\rv{To gain deeper insights into the preferences and considerations of people with SMA when designing VR gestures, we analyzed the participants' feedback and identified four main mental models.}

\sout{There were some overlaps between the gestures designed by different participants. After removing overlaps, we found 233 unique user-defined gestures. In Figure \ref{fig:Distribution}, we can see the different types of gestures used by participants to respond to VR commands. Among these unique gestures, 44.2\% were only-hand gestures, 12.8\% were only-eyes gestures, 5.8\% were only-wrist gestures, 4.5\% were only-forearm gestures, 4.2\% were only-head gestures, 6.1\% were only-mouth gestures, 1.3\% were only-arm gestures, 1.9\% were only-torso gestures, 6.1\% were hand \& wrist, 6.7\% were hand \& forearm, 1.3\% were eyes \& hand, 2.9\% were mouth \& UI and the rates of the other combined gestures were less than 1\%.}

\subsection{Mental Model Observations}
\subsubsection{Use Local Movements to Map Unfeasible Large Range Motion}
From the participants' self-reported abilities, we learned that they have significant difficulty performing large actions, particularly those requiring upper arm strength. These discrepancies between the limitations in upper arm muscle strength and the large motion-scale tasks in VR influenced their experience, leading to self-disappointment and aversion towards VR devices. \sout{as P11 mentioned,"...I cannot reach some targets at the edge of the screen. I got a game over soon and this discouraged me. I feel powerless and disappointed in my strength. People like us are not suited for VR..."
}
\begin{figure*}[t]

    \centering
    \begin{subfigure}[t]{0.3\textwidth}
        \centering
        \includegraphics[width=\linewidth]{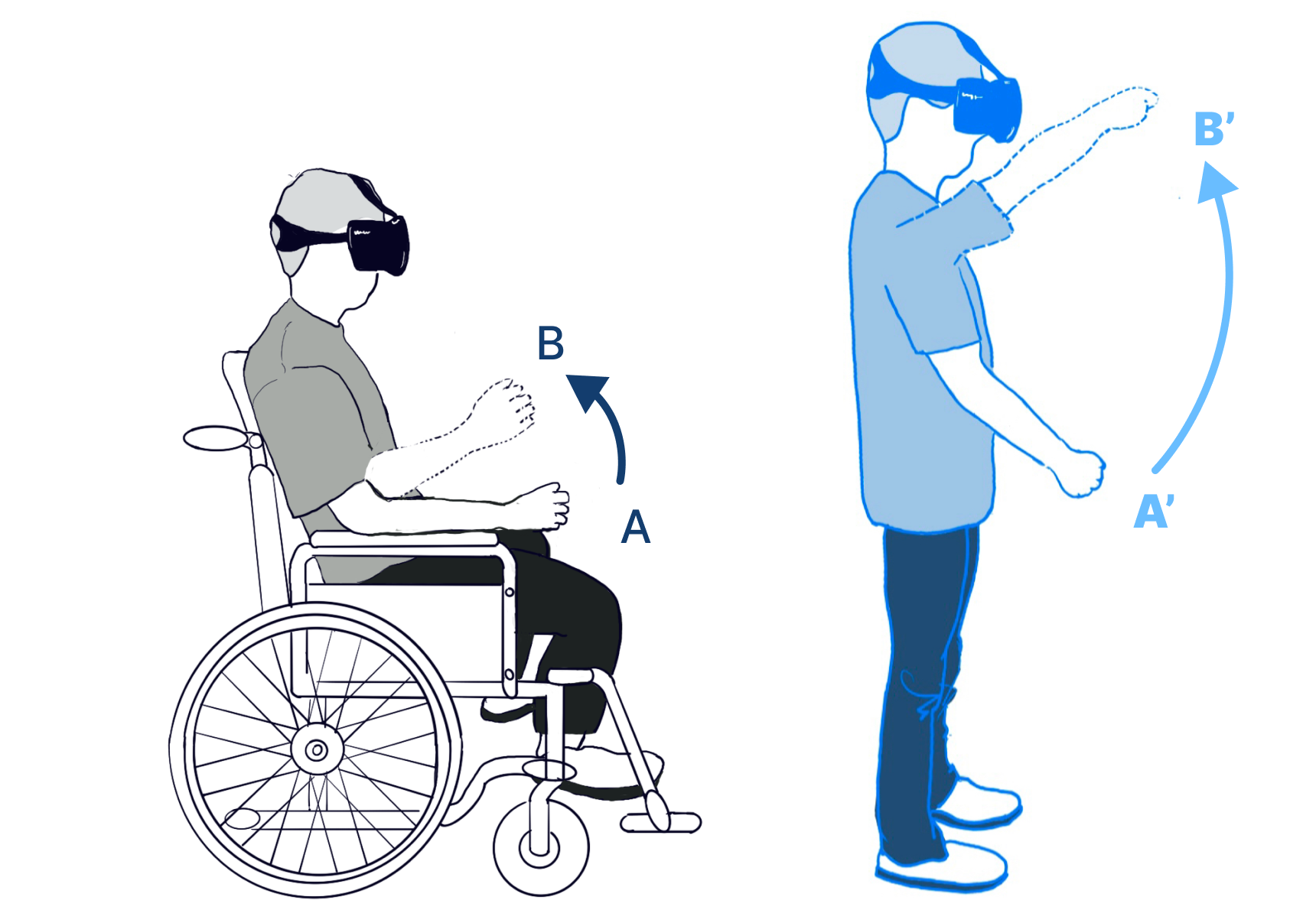}
        \captionsetup{width=.8\linewidth}
        \caption{Participants \rv{prefer to use their hand to make} a small\sout{, energy-efficient} gesture to \rv{map a large movement in VR.}}
    \end{subfigure}%
    \begin{subfigure}[t]{0.3\textwidth}
        \centering
        \includegraphics[width=\linewidth]{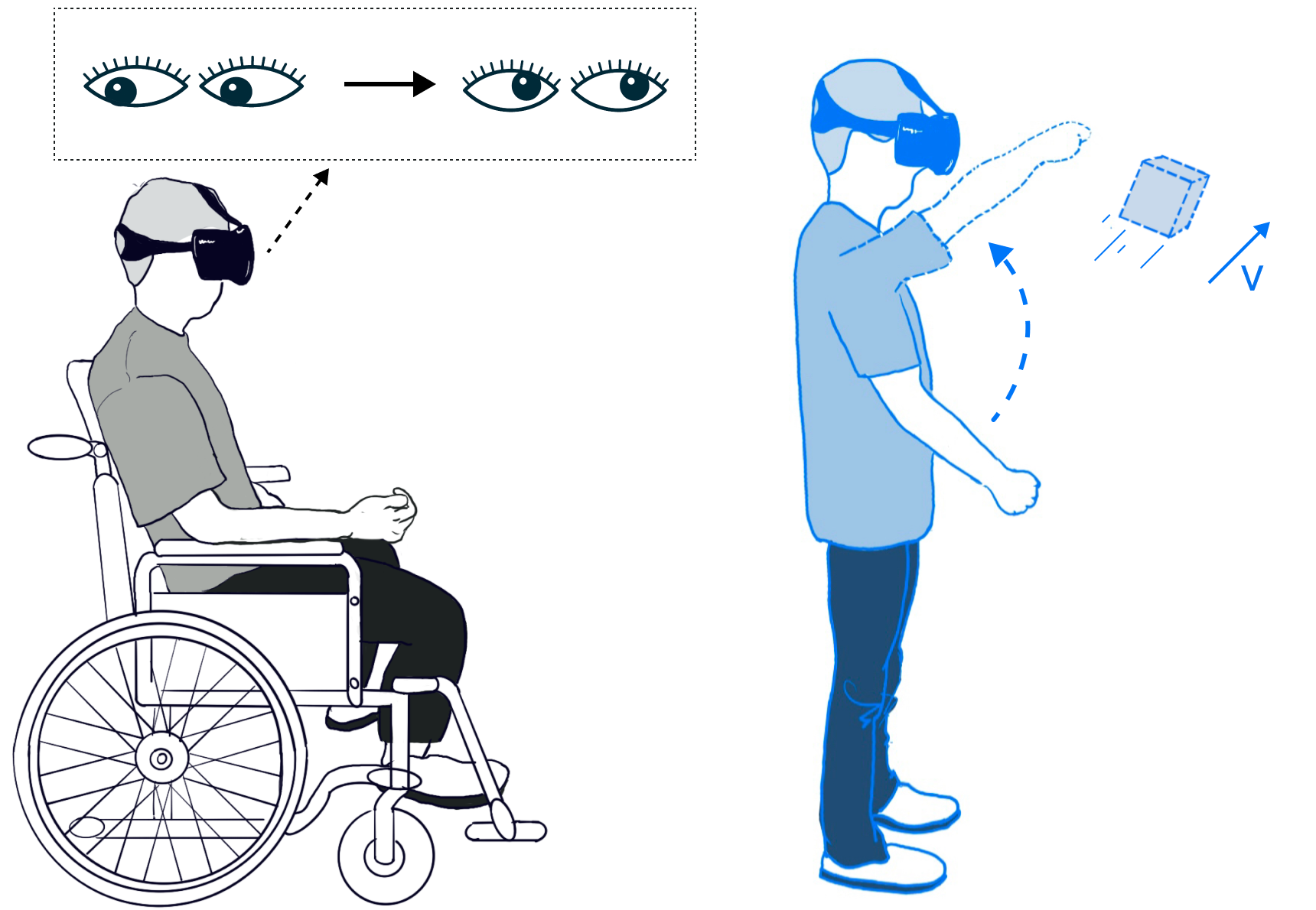}
        \captionsetup{width=.8\linewidth}
        \caption{Participants employ eye movements that mirror the hand gesture required to execute VR commands, such as ‘throw 3D objects.’}
    \end{subfigure}\par\medskip
    \begin{subfigure}[t]{0.3\textwidth}
        \centering
        \includegraphics[width=\linewidth]{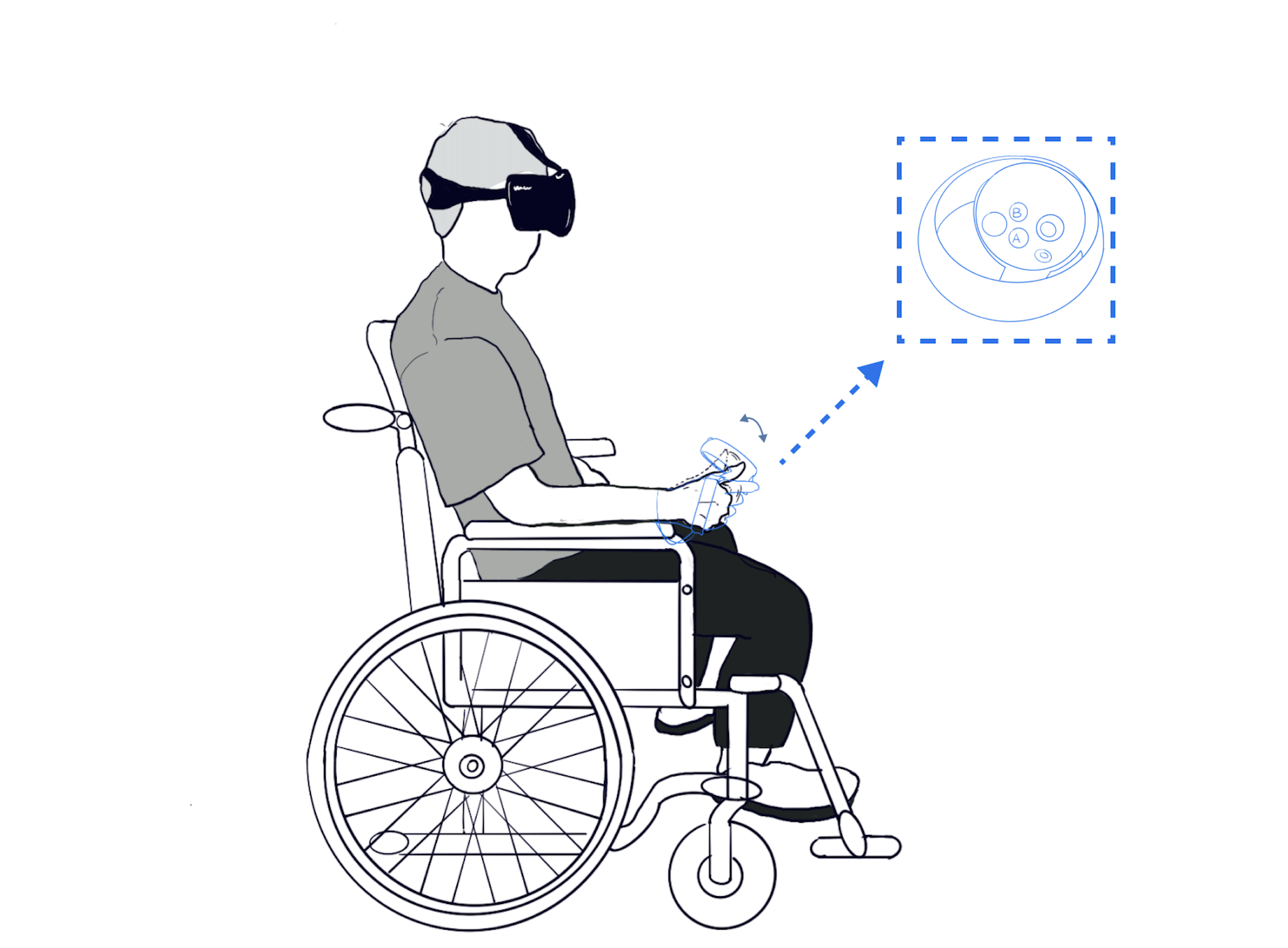}
        \captionsetup{width=.8\linewidth}
        \caption{Participants mimic the use of a VR controller through their gestures.}
    \end{subfigure}%
    \begin{subfigure}[t]{0.3\textwidth}
        \centering
        \includegraphics[width=\linewidth]{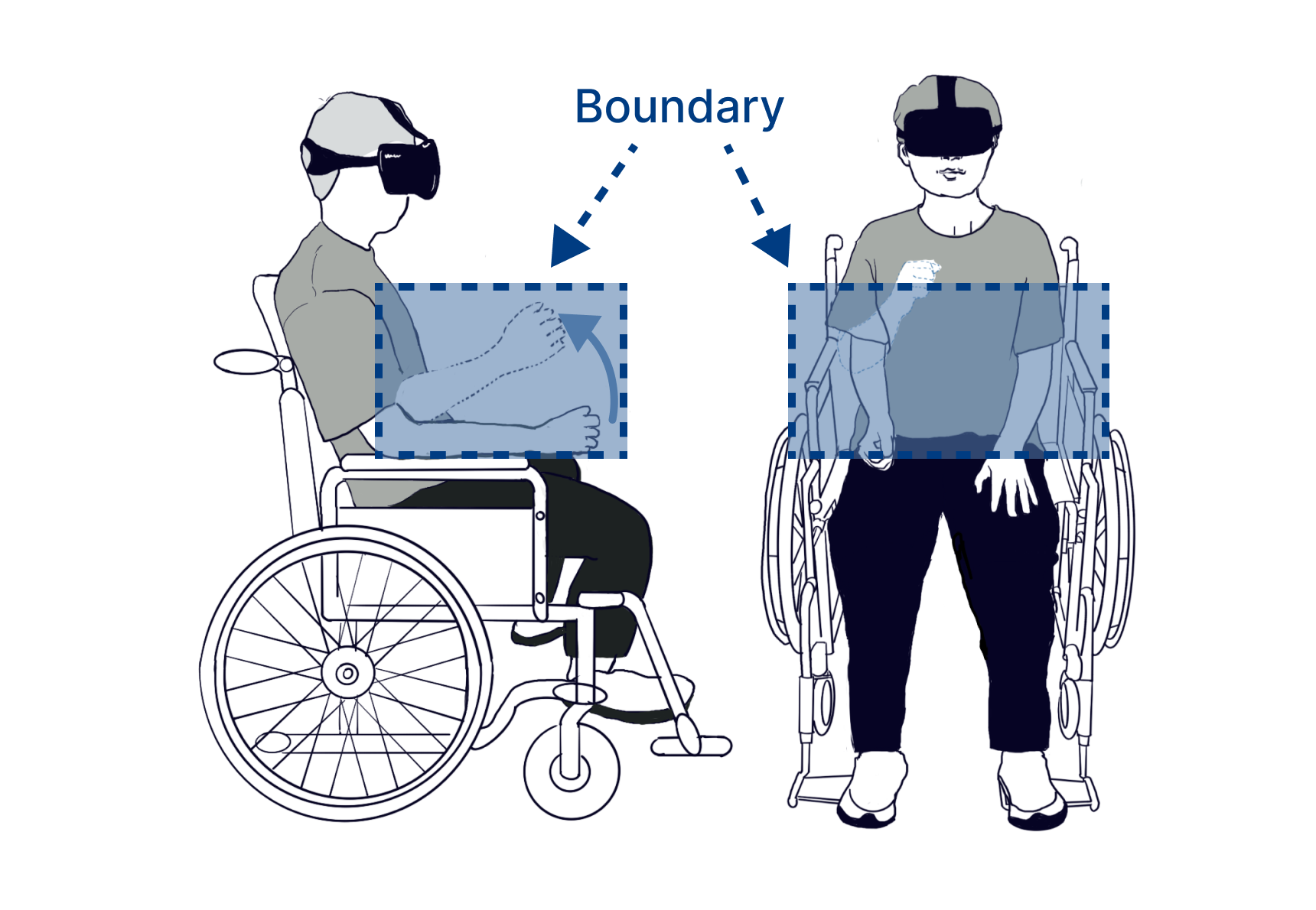}
        \captionsetup{width=.8\linewidth}
        \caption{Participants adjust the height of their gestures to fall within a range from the waist to just below the chest.}
    \end{subfigure}%
    \begin{subfigure}[t]{0.3\textwidth}
    \centering
        \includegraphics[width=\linewidth]{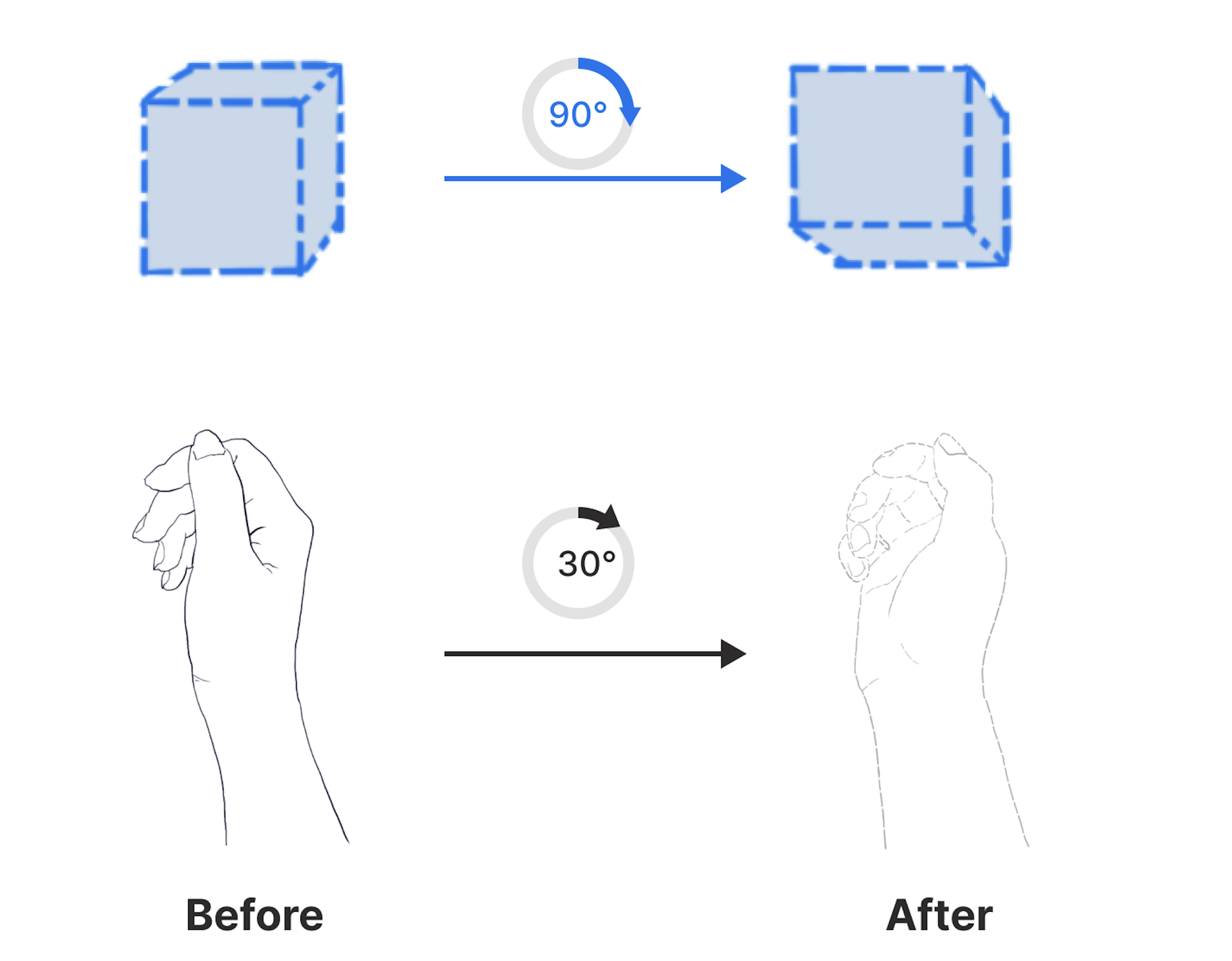}
        \captionsetup{width=.8\linewidth}
        \caption{Participants rotate their wrists by a smaller degree, such as 30°, to manipulate the virtual object by a larger degree, such as 90°.}
    \end{subfigure}
    
    \caption{\textbf{The example illustration of five strategies for conducting large actions in VR: (a) Utilize Retaining Hand Mobility to Mimic Weak Proximal Upper Limbs; (b) Substitute with above-the-neck body parts; (c) Imitating Existing Peripherals; (d) Restrict the performance boundary of gestures; (e) Enhance small movements through offset.}}
    \label{fig: Strategies Illustrations}
    \Description{Figure 10 shows the example illustration of five strategies for conducting large actions in VR: (a) Utilize Retaining Hand Mobility to Mimic Weak Proximal Upper Limbs: The participant makes a small energy-efficient gesture instead of having the avatar perform a large, energy-intensive motion; (b) Substitute with above-the-neck body parts: The participant utilizes wide-rang eye movements while crafting a gesture for the VR command "throw 3D objects"; (c) Imitating Existing Peripherals: The participant mimics the use of VR controller through their gestures; (d) Restrict the performance boundary of gestures: The participant adjusts the gesture height within the range from below the chest to the waist; (e) Enhance small movements through offset: The participant rotates the wrist by 30° to manipulate the virtual object by 90°.}
\end{figure*}

\textbf{Strategies for Conducting Large Actions in VR.} We identified five strategies that participants employed to complete the large-scale actions while also preserving their body involvement in VR.

1) \textit{Utilize Retaining Hand Mobility to Mimic Weak Proximal Upper Limbs.} Participants (N=10) with remaining hand abilities initially opted for an approach that involved using their hands to mimic the weaker proximal upper limbs, as shown in Fig \ref{fig: Strategies Illustrations} (a). Although other distal body parts may exhibit superior mobility compared to the hands, participants demonstrated a preference for utilizing their remaining hand mobility to enhance body involvement in VR, as P9 noted, {\itshape "the sense of bodily involvement that arises from using the hands can compensate for the difficulty of performing gestures."} P6 incorporated the wrist and fingers as a substitute for the entire arm in gesture design and humorously described this method as {\itshape" recreating a miniature VR world with fingers."}\sout{, as she explained.}

In addition to using their hands, participants also expanded their hand mobility with other tools. For example, when designing a chopping action for 3D objects, P10 picked up a pen as an aid to extend her hands, allowing the limbs below the wrist to become a complete arm.

2) \textit{Substitute with above-the-neck body parts.} Participants with severe atrophy of hand muscles, like those with SMA-I, tend to rely on using body parts above the neck to substitute small movements for larger ones. Their choices of body parts are purposeful. They tend to select body parts with characteristics similar to VR task effects, such as those that can simulate the motion trajectory of 3D objects or the direction of hand force. For example, P3 chose to move her eyes from the bottom left corner to the upper right corner when designing a gesture for VR command \textit{throw 3D objects}, as shown in Fig \ref{fig: Strategies Illustrations} (b).

3) \textit{Imitating Existing Peripherals.} Participants used their body parts to imitate existing devices to accomplish large movements. For example, when designing locomotion tasks, participants found it challenging to design gestures for tasks with a sense of distance. When unable to think of a direct body part to use, they referred to existing devices capable of achieving locomotion. P6 and P11 used their wrists and thumbs, respectively, as analogies for the joystick, as shown in Fig \ref{fig: Strategies Illustrations} (c).

4) \textit{Restrict the performance boundary of gestures.} All participants expressed their desire to restrict the performance boundary of VR operations. Participants with SMA Type I or Type II, who primarily adopt a lying position, wished to limit the range around the waist. For participants with SMA Type III, they hoped to control the height within the range from below the chest to the waist, as shown in Fig \ref{fig: Strategies Illustrations} (d).

5) \textit{Enhance small movements through offset.} After designing micro gestures, participants also sought to enhance the effects of small movements through offset, as shown in Fig \ref{fig: Strategies Illustrations} (e), including using discrete gestures to initiate continuous actions and using small gestures to large movements. For example, P10 used wrist rotation to rotate 3D objects, but she chose to {\itshape"rotating the wrist by 5 degrees in reality, which would correspond to rotating the virtual object by 25 degrees."} For the same task, P11 hoped to activate continuous rotation by slightly rotating her wrist.

\textbf{Concerns about the Accuracy of Recognition of Micro Gestures.} Although we had reminded the participants before the design phase that there was no need to consider whether the current technology or equipment could support their gesture design, participants still worried about  whether the VR device could accurately recognize their micro gestures, as P2 said, {\itshape "My mouth can only open as wide as one finger can fit in, and I am not sure if the device can recognize it."} Besides concerns about micro-gesture non-recognition, participants also worried that limiting the range of gesture performance might cause gestures to fall outside the recognition area. 

In addition to being unrecognizable, participants also expressed concerns about gesture misrecognition due to limb tremors, particularly when the movement range is relatively subtle. Hand tremors are a common condition among our participants. P8’s physical ability is better than other participants, but his hand tremors still significantly impact his daily life, such as mistyping.

This concern about the accuracy of gesture recognition affects the participants' perception of the design process's difficulty and satisfaction with the designed gestures. However, participants still hope to rely on \rv{the improvement of VR performance to reduce their burden.} \sout{ VR device, as P5 commented, "If I change another design, it would be difficult for me to conduct. I would rather leave this issue to the technical staff and not leave this burden for us."}
In addition to expecting high-precision gesture recognition, participants also hope that the VR system recognizes the motion trend rather than the specific moving body parts. For example, when P7 designed the action of waving toward target objects, he mentioned,{\itshape "I hope VR can recognize the direction of the swing, regardless of whether it is performed with the arm or wrist."}

\subsubsection{Minimize Physical Efforts within Capabilities}
Participants would minimize the physical efforts exerted in their designed gestures. This may be related to the generalized muscle weakness among people with SMA, specifically manifested in the weakness of force and the difficulty of maintaining it. \sout{For example, P7 faced rapid fatigue when manipulating a bow and arrow in a VR game. He shared,
"I could aim and release the arrow precisely the first or second time. Later on, I had to release it as soon as I raised my arm, and then my arm would fall directly..."}

\textbf{Strategies to Minimize Physical Efforts.} \sout{We observed that}Participants employed three strategies to reduce their physical efforts.
Firstly, all participants would repeatedly measure the smallest body parts used in their gestures, such as the number of fingers or the extent of clenching a fist. Through this continuous testing, they sought to find the most effort-efficient combination of body parts. For example, when designing a gesture for grabbing a 3D object, P11 changed the gesture from using five fingers to two after repeatedly testing. Secondly, participants also considered the duration of the gesture use and tended to use gestures with shorter durations to reduce physical load. Finally, besides the physical effort required for the gesture itself, participants also considered the operation frequency of VR tasks. And they tend to prioritize using the most flexible body parts to complete more frequently used tasks. For example, participants (N=4) chose blinking as a gesture for high-frequency VR tasks, as P4 said, {\itshape "I prefer to use blinking because it is relatively easy for me, allowing me to complete the task with minimal effort.” }

\textbf{The Balance Between Energy Conservation and Consistency.} Participants (N=7) are willing to bear the additional physical burden to maintain consistency between reality and the virtual world. For example, P11 was willing to use more fingers in the gesture for pulling a 3D object to simulate the force performed in the task. \rv{However,} \sout{Although the participants were willing to endure a greater physical burden for consistency, it was under the condition that they believed that completing more strenuous actions was still within their physical capabilities.} when designing gestures that require significant physical effort, they potentially sacrifice consistency. For example, when designing the gesture for pinching an object, P7 opted to abandon the more experiential loose grip in favor of a more effortless tight grip.

\subsubsection{Consider Social Encouragement and Acceptance.} Participants prefer gestures with a cool physical appearance to convey to others that they are doing something cool and to encourage themselves to use these gestures more frequently. They also try to avoid gestures that may evoke negative associations.

\textbf{Gestures with Cool Physical Appearances.}
Participants tried to design gestures with cool physical appearances based on the characteristics of VR. Participants mainly drew inspiration from popular culture, magical stories, and science fiction films to design appealing gestures. For example, when designing a gesture for hitting 3D objects, P7 chose classic moves from the martial arts world, a Chinese popular culture, because he thought {\itshape "both VR and martial arts can transcend human limitations to accomplish some incredible things"}, and he also believed that using popular ethnic culture could reduce the understanding barriers between VR users and viewers.

 The cool and creative aspects of gestures can increase participants' satisfaction and enhance engagement. Hence, when designing gestures that cannot fully utilize the characteristics of VR, participants' satisfaction may decrease. P8 mentioned that he was not satisfied with his designed gestures and was not very willing to use them due to a lack of creativity. He explained, {\itshape \rv{"}I think VR is a very technologically advanced product, and cool gestures also make VR more appealing."} 

\textbf{Social Acceptance.} Participants were concerned about others' perceptions when designing gestures. They were reluctant to design gestures that were too unusual or would evoke negative associations. In addition to negative associations, P7 mentioned the unique understanding barriers caused by the isolation of VR users from others.

\subsubsection{Design Gestures across Time Span and Abilities.}\rv{Participants design gestures} not only based on their\sout{limited} abilities but also \rv{consider} their past experiences and potential future physical conditions. 

On the one hand, participants would reflect on their previous better physical conditions or even refer to people without motor impairments, and they tend to design gestures that resemble those performed by individuals without motor impairments. This approach allows them to gain the perception that their motor abilities remain intact. As P4 expressed\sout{ her willingness to take on a greater physical burden to design gestures that closely align with a physical fitness state, as she stated},
{\begin{quote} 
"When designing a gesture, I think about how I would perform the tasks if I did not have a disability, and then I try to get as close to that state as possible. It is like an idealized state where I imagine myself being the same as before, or even without a disability. In this way, I can do anything in VR, just like able-bodied people."
\end{quote}}
On the other hand, participants also expressed concerns about progressive muscle atrophy and would like to design a universal gesture set for the future. For example, at the end of the design process, P2 proposed to redesign all gestures by combining UI components with body parts, as he said,
{\begin{quote}
"If possible, I would list all operations on the side of the screen, and I could either look at it for a few seconds or tap my nose to confirm."
\end{quote}}
P1 believed that having a backup solution was necessary and could provide him with a sense of security. However,P1 also acknowledged the lack of immersion and body involvement with this method. \sout{as he said,"The immersion would be worse, so I thought of it but did not choose this method even when I can not find suitable body parts [for gestures]. But if more of my body parts can not be used in the future, having a backup plan like this will make me feel at ease."}

%% file: Discussion.tex

In this paper, we discuss the implications of the user-defined upper-body VR gestures designed by and for people with SMA. VR offers an opportunity for individuals with limited mobility to act beyond their physical capabilities, promoting inclusivity and equality \cite{gerlingVirtualRealityGames2020,lewisAssistiveTechnologyLearning1998}. However, current VR devices with ability assumptions pose challenges in input methods for them \cite{mottJustWentIt2020,kouroupetroglouAssistiveTechnologiesComputer2014a}, and there are few attempts made at alternative accessible VR input methods for users with motor impairments \cite{twoinone}. \rv{This study} \sout{firstly conducted a video content analysis to gather VR commands. Then we presented the effects of commands to participants in}\rv{used} a video elicitation study to investigate what upper-body gestures people with SMA prefer in VR. This was pivotal for us to understand their design considerations and to provide some useful design suggestions.
\subsection{Key Takeaway}
By involving people with SMA in the design process, we identified a taxonomy of user-defined upper-body gestures and their mental models. All participants in the elicitation study expressed their desire to experience VR with gestures in the future. \sout{The hands are the most frequent body part they used when designing VR gestures.} Gestures involving hands were the most diverse and preferred. The type of task and participants' abilities influence the choice of body parts for gesture design. \rv{We identified four mental models that people with SMA employed when designing gestures.} They preferred to use local movements to map unfeasible extensive motion, intending to minimize physical effort in their gestures. \rv{They also focused on designing gestures with visually appealing appearances, and they aimed to create gestures adaptable to changes over time and their abilities.}

\subsection{Design Considerations for Accessible Gesture Input in VR}
\sout{Compared to the accessible gesture inputs for users with upper-body motor impairments for smart phone \cite{zhangSmartphoneBasedGazeGesture2017,fanEyelidGesturesMobile2020,fanEyelidGesturesPeople2022} and wearable devices \cite{vatavuUnderstandingGestureInput2022b}, our user-defined gestures are unique in two aspects. Firstly, our gestures are designed for VR. Mott et al. identified seven barriers related to the physical accessibility of VR devices that people with limited mobility might encounter, including inaccessible controller buttons and difficulty in manipulating dual motion controllers \cite{mottJustWentIt2020}. They later proposed an accessible method by mapping unimanual input into bimanual interactions in VR, allowing users with limited mobility to perform bimanual interactions \cite{twoinone}. However, what we proposed is to use user-defined upper-body gestures to achieve VR input accessible for people with motor impairments, grounded in their ability and creativity.

Secondly, our research specifically targets people with Spinal Muscular Atrophy (SMA). In contrast to previous studies that focused on individuals with various limited mobility, we found that people with SMA tend to use local movements to map unfeasible extensive motions when designing gestures, due to their significant difficulty in performing large-scale actions. Moreover, for people with SMA, minimizing physical effort within their capabilities is an essential design consideration, and this result is consistent with previous works about people with motor impairments' preferences to input methods \cite{FreedomtoChoose,SmartwatchInteractions}}

Our findings suggest the need for design approaches that capitalize on users’ motor abilities and preferences. In the following, we present four practical implications informed by our empirical findings to facilitate the development of accessible VR input methods for people with motor impairments.

    \textbf{Design visually appealing gestures to encourage people with motor impairments to engage in VR.} 
\rv{Our findings highlight the importance of creating VR input gestures that are visually appealing to encourage people with motor impairments to engage in VR. Prior studies also identified that social acceptance is a crucial consideration in the design of gesture-based interactions for people with motor impairments in other contexts \cite{fanEyelidGesturesMobile2020,zhaoDonWantPeople2022}. However, our study extends beyond the realm of social acceptance. Informed by our study, the aesthetics and social encouragement address a need that goes beyond mere social acceptance and can enhance the engagement of users with motor impairments.} Therefore, when designing accessible VR input gestures for people with motor impairments, it is essential to fully utilize the characteristics of VR\sout{and VR tasks} to create gestures that are not only functional but also \rv{appealing} and engaging.

  \textbf{Improve recognition accuracy of micro gestures by people with motor impairments in VR.} Participants \rv{expressed concerns} about whether\sout{whether the} VR devices can accurately detect their micro-gestures. \rv{Efforts have been made to improve gesture recognition accuracy \cite{DigiTap,FingerInput,basninIntegratedCNNLSTMModel2021,EyeGazeMicrogestures}}. However, these approaches primarily target input methods designed for able-bodied individuals, potentially leading to technological incompatibility. 
  \rv{Our study's focus on this issue highlights the need for technology that accommodates the specific challenges faced by people with motor impairments, such as limb tremors in SMA, necessitating specialized gesture recognition technologies.}

       \rv{\textbf{Designing more personalized user-defined gestures for people with motor impairments.} While designing gestures for people with SMA, a standard gesture set might not be optimal for a particular user. The results of our gesture analysis indicate that they have different physical conditions and habitual perceptions. And the difference in VR commands may also influence their preference. Thus, it is important that an individual user with motor impairment can customize their VR gestures, similar to the work in the mobile phone context by Ahmetovic et al \cite{ahmetovicRePlayTouchscreenInteraction2021a}.}
       
    \textbf{Combine alternative or adapted input devices with user-defined gestures for people with motor impairments in VR.} 
\rv{Some participants expressed their interest in combining alternative input devices like joysticks and adapted keyboards with VR gestures, particularly for those with SMA who have weaker anti-gravity hand muscle capabilities. Although it remains unclear how to create an ecology that combines input devices with gestures for people with motor impairments in VR \cite{WentzeMultiModal},
our findings underscore this need to enhance the VR experience while preserving the unique preferences and abilities of users with motor impairments.}
\sout{it is indeed important to consider their preferences and abilities while preserving a complete VR experience.}

    \textbf{Using user-defined gestures for motor rehabilitation in VR.} 
Participants expressed their expectations for VR to facilitate rehabilitation by incorporating their more severely affected body parts into gesture design, which could encourage more frequent use of these areas and potentially slow down muscle degeneration. \sout{However, }Existing VR rehabilitation methods primarily involve designing games with task-specific training scenarios, which can be relatively simple and repetitive \cite{cameiraoRehabilitationGamingSystem2009,kimClinicalApplicationVirtual2020,ROSE2018153}.  \rv{Using user-defined gestures for motor rehabilitation in VR suggests a more natural and personalized integration of rehabilitation exercise into VR interactions.}

\subsection{\rv{Reflections on Gesture Design with Able-bodied Movement References and Video }}

\rv{
    Participants were influenced by the able-bodied movements demonstrated in the video, despite our instruction to focus on their own abilities and use the video examples solely for understanding VR commands. Our observations and interviews revealed that all participants, regardless of their motor abilities, initially perceived the video actions as reference gestures and then adapted them based on their actual motor skills. For instance, when designing the \textit{Pinch a Nearby Object} commands, individuals with relatively stronger motor abilities, such as P12 and P10, crafted hand gestures by incorporating the example movements of pinching with the thumb and forefinger. Conversely, participants with weaker motor abilities, like P1-5, primarily considered their capabilities in the example movements before making adjustments to their designs. According to their feedback, the use of able-bodied movement references not only reduces their memory load but also facilitates the creation of memorable gestures. As P3 mentioned, 

    \begin{quote}
    
      "I study VR gestures used by able-bodied individuals both in daily life and online videos. If there are 100 common gestures, and I create 100 new ones, the total to remember would be 200. However, with some overlap, it could be reduced to 150."
    
    \end{quote}
}

\rv{
    Therefore, as individuals with SMA commonly look to able-bodied movements before designing gestures, we suppose that even when using a VR headset to experience commands firsthand instead of relying on our examples, they still consider the actions of those without impairments as a reference for their gesture design. This can be confirmed in future research, potentially influencing the focus of future HCI designs for individuals with motor impairments.
}

\rv{
In our study, we utilized 2D videos to demonstrate VR command effects within a 3D immersive environment. While this method may lack the complete immersive VR experience, it could potentially limit participants' comprehension of each VR command. For example, P11 and P12 both had queries about differentiating between \textit{Confirm a Nearby Selection} and \textit{Confirm a Far Selection} because they seemed similar in the 2D video, while we addressed these queries through video review and detailed explanations.}

\rv{
To enhance understanding, we engaged participants with prior VR experience and provided verbal descriptions as a supplement for each VR command. However, it remains unknown if participants would change their design in our work when they wore a VR headset and were shown the VR command effects in immersive scenarios.}

\rv{
Future research could explore a more hands-on approach, involving participants with motor impairments in designing gestures directly within a VR environment. Techniques like the ‘Wizard of Oz' could be employed to effectively link gestures with VR command effects. Such an approach has the potential to offer a more authentic and immersive experience, leading to more intuitive and effective gesture design.
}

%% file: Limitation_and_Future_Work.tex
\rv{{\bfseries Impact of Single Camera on Gesture Observation.} Initially, we proposed using two cameras to provide a comprehensive view of participants' gesture design process—one in front and one on the side. However, due to mobility and device constraints, participants were limited to a single camera. This limitation may have hindered our ability to fully observe gesture articulations despite detailed inquiries. When comparing the gestures of two offline participants with those participating online, it appeared that the impact of the scope limitation was minimal. Future studies could enhance insights by integrating offline user studies for comprehensive comparison and validation of findings.}

\sout{{\bfseries Online and Offline Settings.} Due to the geographical dispersion of participants, most interviews were conducted online. Although there was no significant difference between the online and offline participants in the gesture design process and results, we may have missed some details online. Future work should include more in-person studies to obtain more comprehensive details. }

{\bfseries Limited Coverage of People with SMA.} 
\rv{ Another limitation of this study is the limited number of participants, with a majority being individuals with SMA type III. Due to geographical dispersion, mobility constraints, and privacy concerns of the participants, we chose to conduct the experiments via online video. To ensure participants can understand the VR referent, we tried to recruit participants with prior VR experience, though the current inaccessibility of VR input methods poses challenges for individuals with SMA types I and II to use VR. Considering the diversity in motor abilities across SMA types, our study ultimately involved 12 participants (2 SMA-I, 3 SMA-II, and 7 SMA-III). However, our participants exhibited the diversity in their gesture design process and results. Future research could include a broader spectrum of participants, particularly focusing on those with SMA types I and II, to gain deeper insights into their considerations.}

\sout{Individuals with SMA types I and II have more limited spontaneous motility, making it challenging for them to experience current VR devices. Since the participants with SMA types I and II in our study were experienced gamers and thus basically have no issues in understanding VR tasks with referents.} 

\sout{{\bfseries Lack of an Experiential System.} We did not develop a prototype system for participants to directly experience the effects of their designed gestures in VR, such as using the "Wizard of Oz" technique to enable participants to connect their gestures with outcomes. Instead, participants could only infer their preferences based on the video elicitation and their past VR experiences. The user preferences obtained in this study can serve as a reference for future work that aims to develop a system to use user-defined gestures to improve accessibility for people with impairments in VR.}

{\bfseries Potential Impact of Non-exhaustive Command Selection.} \rv{We identified 26 common commands from the popular applications from three VR application categories. This strategy was employed in an effort to capture a diverse and representative set of common VR commands. However, we acknowledge this approach may not fully capture the wide range of interaction possibilities within VR and tends to focus more on current, prevalent technologies. This limitation might restrict the breadth of our findings. For instance, the number of commands identified in different categories may vary, potentially affecting the distribution of body parts used in each command category. Future research could expand the range of command selection, exploring a wider array of VR applications, including those that are emerging. And the common commands we categorized can serve as a foundational reference for further research.}

%% file: Conclusion.tex
We have adopted a user-centered approach by involving participants with SMA to design user-defined upper-body gestures for VR interactions. We initially identified 26 common commands in VR through a content analysis of 60 videos and then derived a taxonomy of user-defined gestures by analyzing the 312 gestures. Gestures involving hands were the most diverse and preferred. The type of task and participants' abilities influence the choice of body parts for gesture design. The participants preferred using localized movements to map unfeasible extensive motions, aiming to engage their hands for better body involvement despite their limited hand mobility.  They favored gestures that required minimal physical effort based on their abilities and had visually appealing physical appearances based on their perception of VR characteristics. Additionally, they aimed to create gestures adaptable to changes over time and their abilities. In light of these findings, we highlight design considerations and demonstrate future work to enhance the accessibility of VR for people with motor impairments.